\documentclass[preprint,amsmath,amssymb,aps]{revtex4}
\usepackage{graphicx}
\usepackage{dcolumn}
\usepackage{bm}
\usepackage{graphicx}
\usepackage{amsmath}
\usepackage{epstopdf}
\usepackage{amssymb}
\usepackage[export]{adjustbox}
\usepackage{commath}
\usepackage{eqnarray,amsmath}
\usepackage{gensymb}
\usepackage{soul,xcolor}

\renewcommand{\vec}[1]{\mathbf{#1}}
\usepackage{color}
\begin{document}
\title{Flagellar flows around bacterial swarms}
\author{Justas Dauparas}
\author{Eric Lauga}
\email{e.lauga@damtp.cam.ac.uk}
\affiliation{Department of Applied Mathematics and Theoretical Physics, University of Cambridge, CB3 0WA, United Kingdom}
\date{\today}

\setstcolor{blue}
 
\begin{abstract}
 
Flagellated bacteria on nutrient-rich substrates can differentiate into a swarming state and move in dense swarms across surfaces. A recent experiment  measured the flow in the fluid around  an \textit{Escherichia coli} swarm  (Wu,  Hosu and  Berg, 2011 Proc.~Natl.~Acad.~Sci.~USA {\bf 108} 4147). A systematic chiral flow was observed in the clockwise direction (when viewed from above) ahead of the swarm with flow speeds of about $10~\mu$m/s, about 3 times greater than the  radial velocity at the edge of the swarm. The working hypothesis is that this flow is due to the action of  cells stalled at the edge of a colony that extend their flagellar filaments outwards, moving fluid over the virgin agar.  In this work we quantitatively test his hypothesis.  We first build an analytical model of the flow induced by a single flagellum in a thin film and then use the model, and its extension to multiple flagella, to compare with experimental measurements. The results we obtain are in agreement with the flagellar hypothesis. The model provides further quantitative  insight into the flagella orientations and their spatial distributions  as well as the tangential speed profile. In particular, the model suggests that flagella are on average pointing radially out of the swarm and are not wrapped tangentially.
\end{abstract}

\maketitle

\section{Introduction}
Bacteria move by a range of different mechanisms, including swimming \cite{turner2000real}, swarming \cite{turner2010visualization}, twitching \cite{burrows2012pseudomonas}, gliding \cite{mauriello2010gliding}, or sliding \cite{murray2008pseudomonas}, and they can also  form sessile communities attached to a surface, called biofilms \cite{verstraeten2008living}.  In many situations, the presence of a surrounding fluid plays an important role, and appreciable flows  can be observed.  

\begin{figure}[t]
        \includegraphics[width=0.7\textwidth]{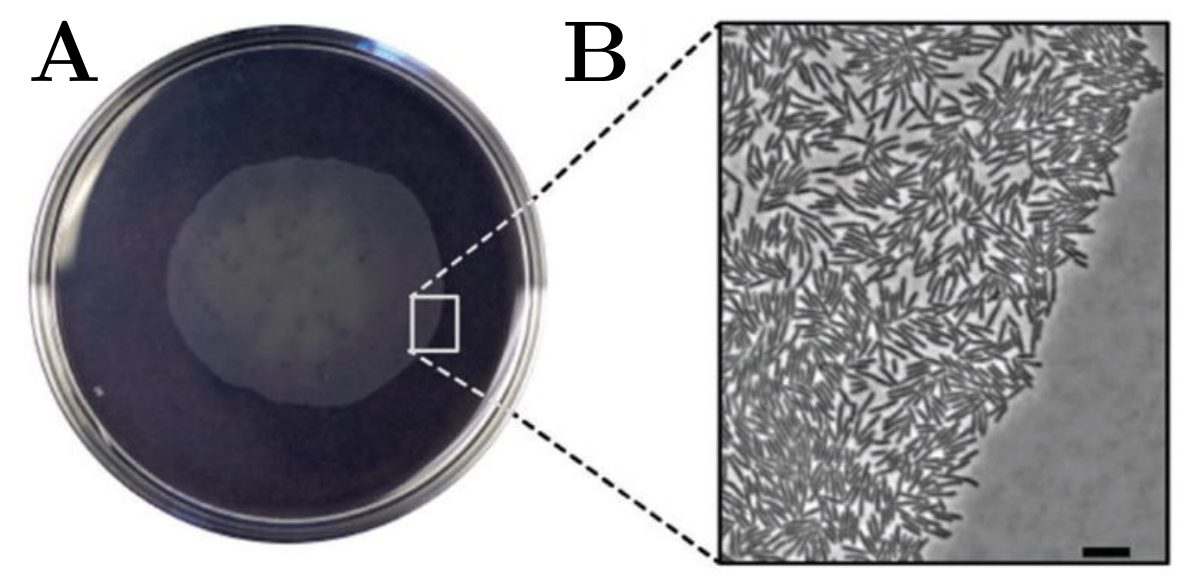}
        \caption{(A) Top view image of a swarming colony of \textit{E.~coli} on the surface of a soft agar gel expanding radially out; the diameter of the Petri dish is 10~cm. 
        (B) Higher magnification images depicting the advancing edge of the \textit{E.~coli} colony; scale bar $= 10~\mu$m. Reproduced from Ref.~\cite{swiecicki2013swimming} with permission of The Royal Society of Chemistry.}
    \label{fig:1}
\end{figure}

Recent flow measurements have shed new light on   bacterial swarms, which are the focus of the current paper. 
During  swarming motile populations  rapidly advance on moist surfaces (see pictures of  a swarm in Fig.~\ref{fig:1}).
In order to swarm,   cells must be motile with functional flagella,  be in contact with, or close to, surfaces and with other motile cells \cite{copeland2009bacterial}. Furthermore, bacteria have to reach a certain cell number before the process is initiated, a feature known as the swarming lag \cite{kearns2010field}.  When cells are grown on a moist nutrient-rich surface, they differentiate from a vegetative to a swarm state. They elongate, make more flagella, secrete wetting agents, and move across the surface in coordinated packs. Many bacteria produce surfactants as they swarm which influence the patterns of expansion  \cite{harshey2003bacterial, berg2005swarming}. However, these surfactants are not essential; in particular, there is no indication that the model bacterium 
{\it Escherichia coli} (or \textit{E.~coli}) produces surfactants \cite{berg2008coli, blount2015unexhausted}. 

There has been considerable progress in understanding  swarming, including cell elongation, increased flagellar density, secretion of wetting agents, increased antibiotic resistance and suppression of chemotaxis \cite{benisty2015antibiotic, copeland2010studying, tuson2013flagellum, partridge2013swarming}. Bacterial swarms display large-scale swirling and streaming motions inside the swarm and it has been recently suggested that swarming bacteria migrate by L\'evy walks \cite{ariel2015swarming}. Moreover, bacterial swarms can take
different appearances but the significance of any particular pattern is unclear and patterns change depending on the environment \cite{ingham2008swarming, ben2006self, wu2015collective}. For the swarm to move, it is important to overcome a number of obstacles, including drawing water to the surface from the agar beneath, overcoming the frictional surface forces and  reducing  surface tension \cite{partridge2013swarming}. Swarms are usually organized with multiple layers of cells in the middle and a cell monolayer near the edge where the fluid film height is about 
the diameter of the cell, i.e.~$\approx1~\mu$m.

\begin{figure}[t]
        \includegraphics[width=0.7\textwidth]{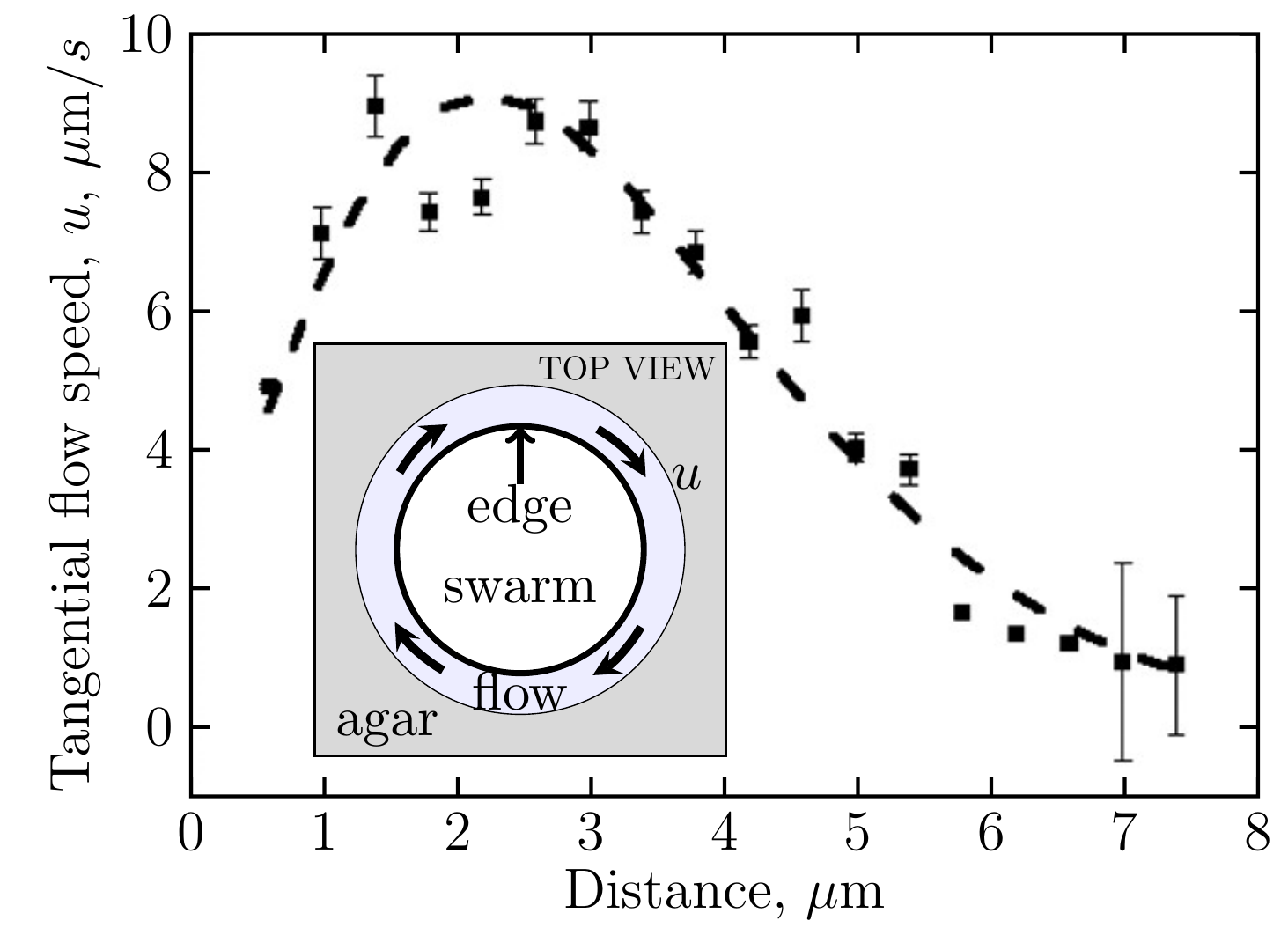}
        \caption{Experimental measurements of chiral flow reproduced from Ref.~\cite{wu2011microbubbles} with permission: Average tangential flow speed (in $\mu$m/s) as a function of the distance from the swarm edge (in $\mu$m), computed from 28 passively-advected bubble trajectories. Error bars are standard deviations for the ensemble of bubble trajectories. 
        Inset: schematic picture showing a top view of a swarm with chiral flow around it.}
    \label{fig:2}
\end{figure}

A recent study addressed experimentally the nature of the flow ahead of an
\textit{E.~coli} swarm \cite{wu2011microbubbles}.  Analysis of small air bubbles used as passive tracer particles  showed that there is always a   stream of fluid flowing in a clockwise direction (when viewed from above ) ahead of the  swarm at a rate of order 
$10~\mu$m/s,  about $3$ times greater than the rate at which   the swarm advances.  The experimental measurement of the tangential flow speed as a function of the distance from the swarm edge is shown in Fig.~\ref{fig:2}. This breaking of symmetry, i.e.~the fact that the flow is in the clockwise direction rather than counterclockwise, might be
related to the  counterclockwise rotation (when viewed from behind the cell) of left-handed (LH) helical flagellar filaments.  The chiral flow provides an avenue for long-range communication in the swarming colony, ideally suited for secretory vesicles that diffuse poorly. Furthermore, understanding this physical phenomenon might have implications for the engineering of
bacterial-driven microfluidic devices \cite{gao2015using}.

Further experiments provide additional physical  insight. 
The deposition of MgO smoke particles on the top surface of an \textit{E.~coli} swarm near its advancing edge 
revealed that the top swarm/air surface is in fact stationary \cite{zhang2010upper}. 
Bacteria swarm thus between two fixed surfaces, a surfactant monolayer above and an agar matrix below. This was further confirmed by noting that
the spreading rates were the same in the case of a swarm spreading in air and spreading under a sheet of PDMS \cite{turner2010visualization}. However, a mobile-super diffusive upper surface has been observed for \textit{Serratia marcescens} and \textit{Bacillus subtilis} \cite{be2011collective}.
Recent measurements  on osmotic pressure in a bacterial swarm \cite{ping2014osmotic} confirmed the drift of air  bubbles (the chiral flow) also suggesting that high osmotic pressure at the leading edge of the swarm takes water from the underlying agar and
boosts bacterial motility.

\begin{figure}[t]
        \includegraphics[width=0.55\textwidth]{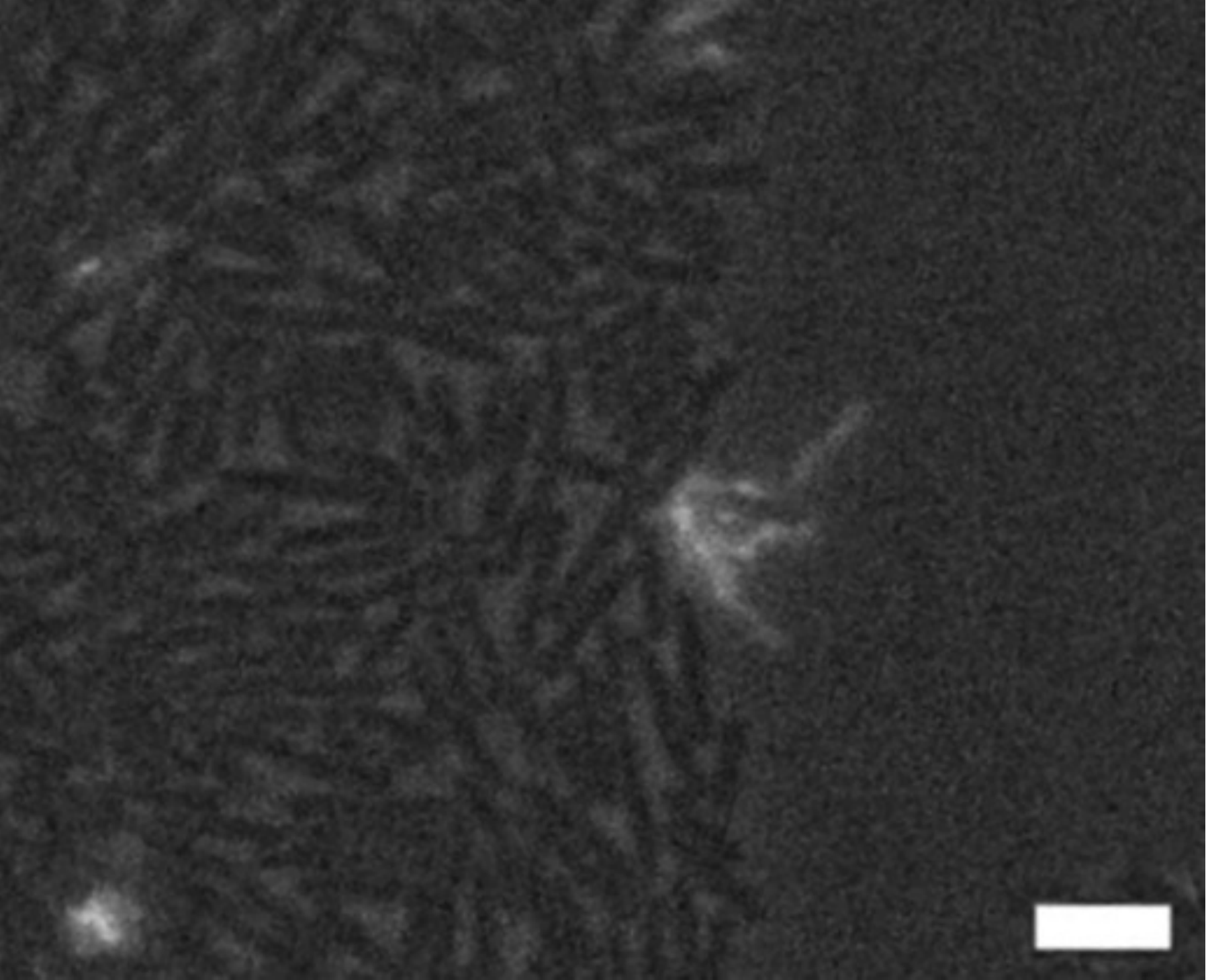}
        \caption{A snapshot from a movie depicting the actuation of flagellar filaments from three bacteria pinned at the leading edge of the swarm; scale bar~$=5~\mu$m. Reproduced from Ref.~\cite{copeland2010studying} with permission from American Society for Microbiology.}
    \label{fig:50}
\end{figure}

At the edge of the swarm, cells rarely get a full body length into virgin agar territory before stalling. When they are stalled at the edge of a colony, cells extend their flagellar filaments outwards, moving fluid over the
virgin agar \cite{darnton2010dynamics}. Using  biarsenical dyes, flagella  have been imaged near the edge of the swarm as well as the extension of liquid film beyond the cells \cite{copeland2010studying} (see a picture of a swarm edge in Fig.~\ref{fig:50}). 
The working hypothesis is that these outward pointing flagella pump fluid outward, contributing to swarm expansion and instantaneously create the observed chiral flow in the clockwise direction. 
A moment later, the swarm expands enough to release the stalled cell, which swims back into the interior or along the swarm edge. In an instantaneous snapshot of the swarm boundary, it appears thus that a ring of non-motile cells lines the edge, but these cells are fully motile once transported back into the  interior of the swarm.

The experimental observations raise outstanding questions, in particular on the detailed nature of the chiral flow around the  swarms. Are the clockwise fluid flows  due to flagella sticking outside the swarms? If yes, are the flows primarily  due to  the rotation of the  flagellar filaments  (i.e.~similar to what would be obtained if all flagella were rigid cylinder radially extending away from the swarm edge) combined with a  breaking of symmetry  in the height of the flagella over the surface? Or are the flows primarily due to the  net propulsive forces exerted by the flagella combined with a breaking of symmetry in their orientation away from the radial direction \cite{cisneros08}?

 In this paper, we present an analytical description for the action of  rotating flagella from stuck cells at the edge of the swarm.  Using known flow singularity solutions,  and deriving new ones, we first build an analytical model of the flow induced by a single flagellum in a thin film. We then use the model, and its extension to multiple flagella, to compare with the experimental measurements. The model leads to results that are quantitatively consistent with the experimental observations and provide quantitative  insight into the flagella orientations, their spatial distributions  and the tangential speed profile. In particular, the model suggests that (1) the observed clockwise flows around swarms are generated by bacterial flagella; (2) the flagella  are on average sticking almost radially out of the swarm  which means that the torque they exert on the fluid is the major contributor to the chiral flow; (3) the chiral flow is in a clockwise direction when viewed from above because the right-handed helical flagella are rotating counterclockwise when viewed from outside the swarm looking radially in and these flagella are closer to the agar surface than the fluid/air interface.

\section{Theoretical model}
\subsection{Geometry and boundary conditions}

\begin{figure}[t]
    \centering
	\includegraphics[width=0.5\textwidth]{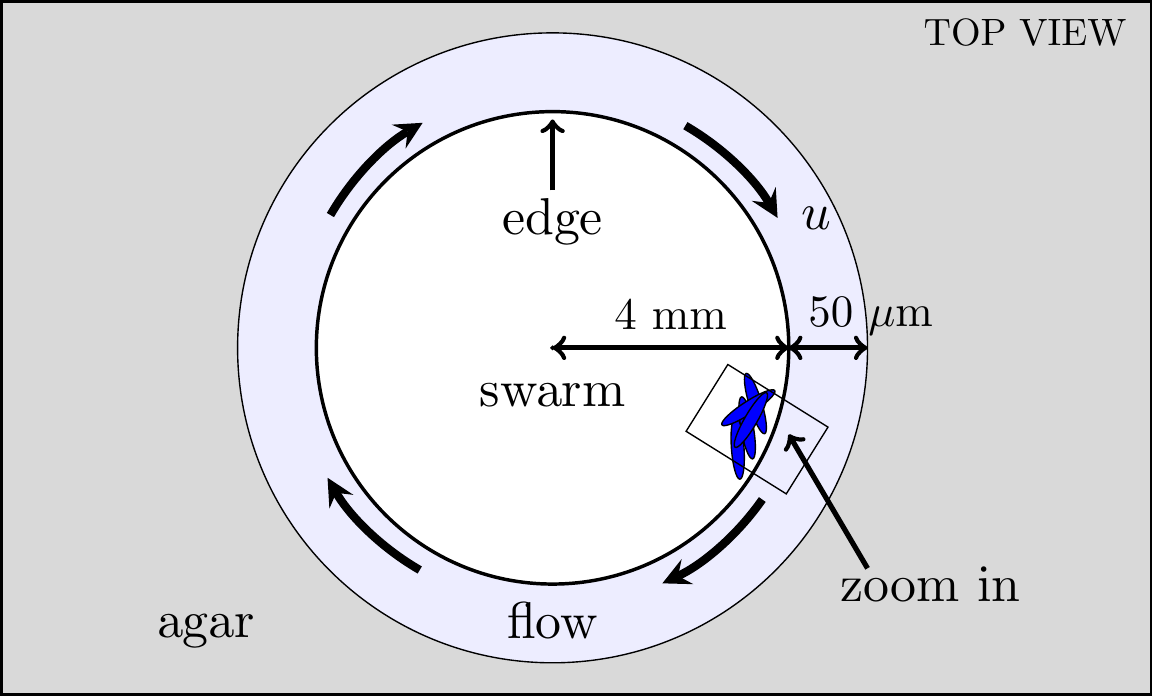}\\
	       \includegraphics[width=0.5\textwidth]{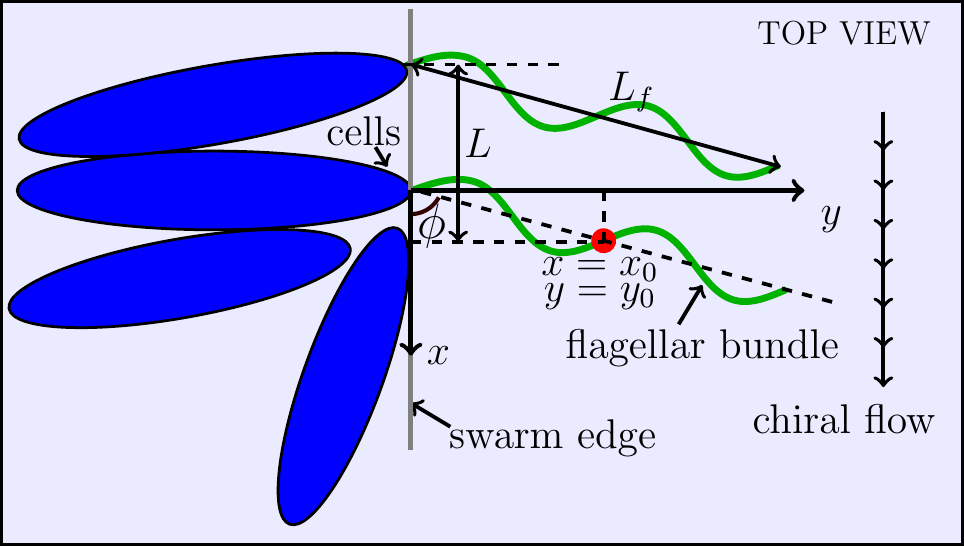}\\
        \includegraphics[width=0.5\textwidth]{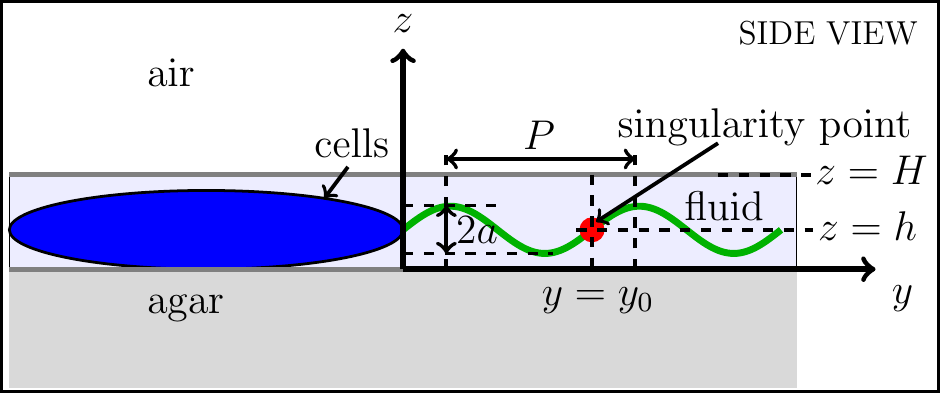}
    \caption{Mathematical model of \textit{E.~coli} cells stuck at the swarm edge above an agar plate ($z=0$) and below a stationary air/fluid interface ($z=H$).      The thickness of the film is  $H=1~\mu$m. The red circle denotes the location of a particular force/torque singularity,  $(x_0,y_0,h)$. 
    The chiral flow is observed in the positive $x$ direction while  the radial flow is in the positive $y$ direction.} 
    \label{fig:3}
\end{figure}

Seeking to  understand  the governing physics behind the chiral flows,  
we use an idealized model with assumptions  based on  experimental observations. First of all, consider a cell  stuck at the edge of the swarm as in  Fig.~\ref{fig:3}.  Each part of a beating helical flagellar bundle (i.e.~an assembly of identical flagella wrapped around each other and rotating in synchrony \cite{berg2008coli}) exerts a force on the fluid. We can calculate these forces using a local drag approximation \cite{lighthill1976flagellar} and take time averages over a rotation of the flagella which result in a net force, $\vec{F}$, acting on the fluid as well as a net torque, $\vec{G}$, acting about the rotational  axis on the fluid. We can either model the instantaneous fluid flow with the helical distribution of forces along the flagellar bundle or we can instead consider  time averages for the fluid flow  and model it with either a single or line distribution of forces/torques along the flagellar bundle -- both will be proposed below. 

Notation for the model is shown in Fig.~\ref{fig:3} (top view, a zoom in of the top view near the swarm edge, and a side view). We use local cartesian coordinates with $x$ tangent to the swarm ($x>0$ being clockwise), $y>0$  in the radial direction and $z$ in the third direction above the swarm. 
Above the cells there is a stationary swarm/air interface, which we assume is  parallel to the agar/fluid interface  and a distance  $H$ away. We thus assume that the fluid satisfies a no-slip boundary condition on the agar surface at $z=0$ as well as  on the swarm/air interface at  $z=H$, consistent with measurements  \cite{wu2011microbubbles}.  Since the radius of the swarm in the $x-y$ plane (about $4$ cm) is much larger than any other length scales,  we can assume that the swarm edge  is locally straight and neglect its curvature. This line of densely-packed cell bodies located at the  swarm edge, i.e.~$y=0$,  prevents most  fluid flow from leaving and entering the swarm, so we apply the no-slip boundary condition there too. Bundles of flagellar filaments   are assumed to  be rotating counterclockwise (CCW) when viewed from outside the swarm looking radially in, and they remain  in the 
in the $x-y$ plane (thus parallel to both upper and lower surfaces). The bundle is a left-handed helix with  pitch $P$,  radius $a$, the axial length $L_f$, while the thickness of the  bundle  is denoted $a_f$. Finally, we denote the distance between the neighbouring bundles $L$ and the angle in the $x-y$ plane between the flagellar bundles and the $x$ axis is designated $\phi$;  small values of $\phi$ (or close to $180^\circ$) will correspond to the situation where the flagella are wrapped tangentially around the swarm edge while $\phi\approx 90^\circ$ is the opposite limit of flagellar aligned along the radial direction.

\subsection{Fundamental flow fields}

The Reynolds number  for the flow generated by the rotating helical flagellar bundle is small, typical $Re = \omega a^2/\nu =O(10^{-7})$. We are thus safely in the steady Stokes regime with linear equations allowing us to 
superpose solutions. The flow between the agar surface and the swarm/air surface can be approximated as the flow between two parallel flat no-slip surfaces.

The modeling approach used in this paper employs  flow singularities in the confined geometry of the thin film immediately outside the swarm edge. We will  use a combination of classical solutions and  newly-derived ones.

\subsubsection{Stokeslet and rotlet between two parallel infinite plates}

The Stokes flow for a stokeslet (point force) between two flat planes was obtained by 
Liron and Mochon \cite{liron1976stokes}
and the flow due to a rotlet by   Hackborn \cite{hackborn1990asymmetric}. Consider   a stokeslet $(F_1,F_2,F_3)$ and a rotlet $(G_1, G_2, G_3)$ located  at $(x_0, y_0, h)$. Denote the planar distance to the singularity location $r=[(x-x_0)^2+(y-y_0)^2]^{1/2}$. In the far field, i.e.~in the limit $r/H \gg 1$, it was shown in these papers that the leading-order term for a parallel stokeslet or rotlet is a two-dimensional source dipole with strength depending parabolically on the distance between the 
two plates. When the stokeslet or rotlet  is perpendicular to the plates, the flow decays exponentially, and similarly for the flow component in the perpendicular direction due to a parallel stokeslet or rotlet.
We non-dimensionalize vertical distances with $H$ and horizontal distances with $k=2\pi/L$. The expression for the far-field flow ($r \gg H$) due to the stokeslet, $\vec{u}^{F}$,  is
\begin{align}
 u_i^{F}(\hat{x} ,\hat{y} ,\hat{z}) &= D_j^{F}(\hat{z})V_{ij}^{SD}(\hat{x} ,\hat{y}),\\
 u_3^{F} &=0,\\
 D_j^{F}(\hat{z}) &=\frac{3}{\mu} F_j H \hat{h}\left(1-\hat{h}\right) \hat{z}  \left(1-\hat{z}\right),\\
 V_{ij}^{SD}(\hat{x} ,\hat{y}) &=\frac{k^2}{2\pi} \left(-\frac{\delta_{ij}}{\hat{r}^2}+\frac{2 \hat{r}_i \hat{r}_j}{\hat{r}^4}\right),
\end{align}
while for the  rotlet, the far-field flow $\vec{u}^{G}$ is
\begin{align}
 u_i^{G}(\hat{x} ,\hat{y} ,\hat{z}) &= D_j^{G}(\hat{z})V_{ij}^{SD}(\hat{x} ,\hat{y}),\\
 u_3^{G} &= 0,\\
 D_j^{G}(\hat{z}) &=\frac{3}{\mu} \epsilon_{3jm}G_m \left(\frac{1}{2}-\hat{h}\right) \hat{z}  \left(1-\hat{z}\right),
\end{align}
where  $\hat{x}=kx$, $\hat{y}=ky$, $\hat{z}=z/H$ and $\hat{h}=h/H$, $\hat{x}_0=kx_0$, $\hat{y}_0=ky_0$, $\hat{r}_i=kr_i$ and $i, j, m=1$ or $2$ (we use the standard notation with 1 and 2 standing respectively for the $x$ and $y$   directions). 

Notably, at leading order both  flows are two-dimensional,  decay spatially as $1/r^2$  and are  parabolic in the vertical direction, 
i.e.~proportional to $\hat{z}(1-\hat{z})$. There is thus no flow at leading order due to a rotlet if $\hat{h}=1/2$, i.e.~if the singularity is placed in the middle between two plates.
Further, we see that  the source dipole due to a stokeslet is in the same  direction as the force applied to the fluid, whereas the source dipole due to a rotlet is in the direction  perpendicular
 to the applied torque. From a dimensional standpoint, we note that the magnitudes of the flows scale as
\begin{align}
u^{F} &\sim F H/(\mu L^2),\\
u^{G} &\sim G/(\mu  L^2).
\end{align}
We can linearly superpose  these two fundamental solutions  as
\begin{align}
 u_i(\hat{x} ,\hat{y} ,\hat{z}) &= u_i^{F}+u_i^{G}= D_j(\hat{z}) V_{ij}^{SD}(\hat{x} ,\hat{y}), \\
 u_3 &= 0,\\
 D_j(\hat{z}) &= D_j^F + D_j^G,
\end{align}
and $i, j=1, 2$.  Note that if the top surface  satisfied no-shear instead of no-slip, we would still be able to apply similar analysis because flow due to a stokeslet or rotlet between no-shear and no-slip parallel plates is also a source dipole to a leading order with different strengths (see Appendix \ref{A} for the derivation of this new solution).

\subsubsection{Source dipole near a no-slip wall}

For this simplified flow, we now can add the effect due to the swarm edge at $y=0$, which we approximate as a third no-slip condition, without running into trouble solving flows in the corner. All we need is to  find the hydrodynamic images for a source dipole next to a wall in two dimensions.

\begin{figure}[t]
\centering
\includegraphics[width=0.49\textwidth]{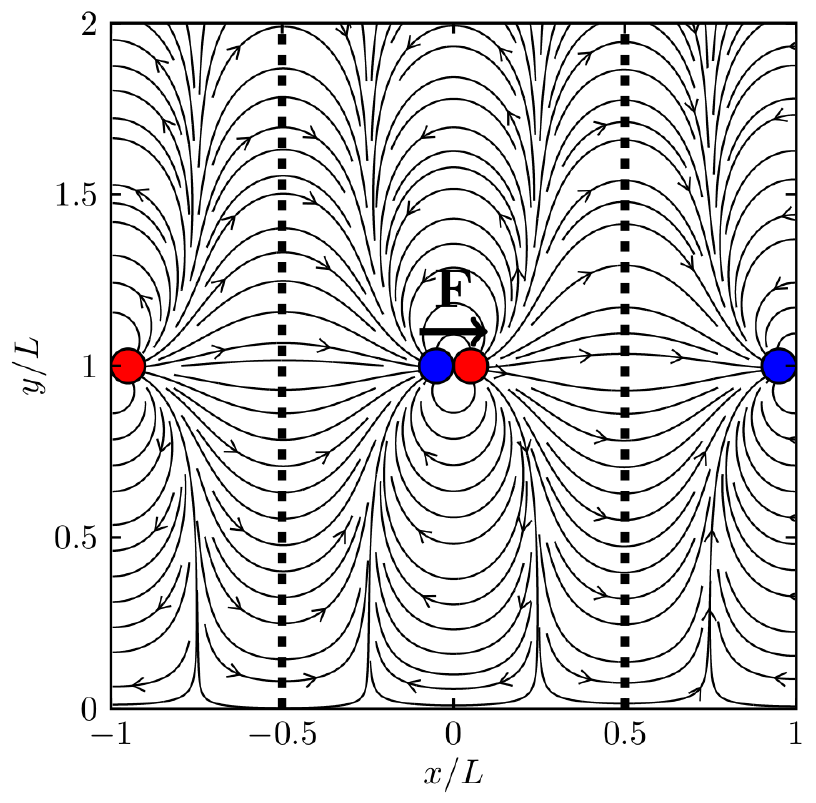}
\includegraphics[width=0.49\textwidth]{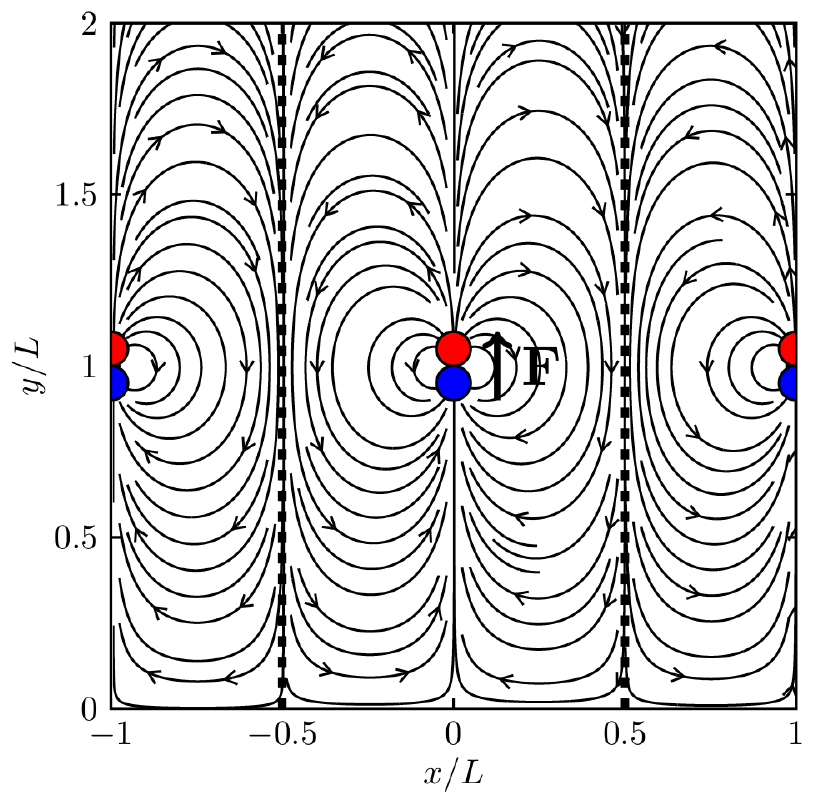}
\caption{Streamlines due to a two-dimensional source dipole parallel (left; $D_1=1$, $D_2=0$) and perpendicular (right; $D_1=0$, $D_2=1$) to a no-slip wall at $y=0$ with periodic boundary conditions along the $x$ direction  (period  $L$). 
Dashed lines are placed midway between singularities showing periodicity.}
\label{fig:4}
\end{figure}

In our case, we have a number of cells along the edge of the swarm and thus a number of helical bundles. We assume that all bundles have the same characteristics and orientation and consider thus a period array of source dipoles at locations $x_n=x_0+nL$, $y_n=y_0$ for $n$ integers where $L$ is the separation between neighbouring
flagellar bundles (i.e.~the period). The flow field for the parallel source dipole of strength $D_1$ and the perpendicular source dipole of strength $D_2$  near a wall is derived in Appendix \ref{B}. Using results from Appendix \ref{B}, we then derive the flow field solutions with periodic boundary conditions  in  Appendix \ref{C}. The obtained flow components are
\begin{eqnarray}
 u_1&=&\frac{D_1 k^2}{2 \pi} \left(\frac{\partial^2 A}{\partial \hat{y}^2} -\frac{\partial^2 B}{\partial \hat{y}^2} - 2\hat{y} \frac{\partial^3 B}{\partial \hat{y}^3} \right)\\
 &&+\frac{D_2 k^2}{2 \pi} \left(-\frac{\partial^2 A}{\partial \hat{x}\partial \hat{y}} - \frac{\partial^2 B}{\partial \hat{x}\partial \hat{y}} - 2\hat{y} \frac{\partial^3 B}{\partial \hat{x}\partial \hat{y}^2}\right), 
 \notag
 \end{eqnarray}
and
 \begin{eqnarray}
  u_2&=&\frac{D_1 k^2}{2 \pi} \left(-\frac{\partial^2 A}{\partial \hat{x}\partial \hat{y}} - \frac{\partial^2 B}{\partial \hat{x}\partial \hat{y}} + 2\hat{y} \frac{\partial^3 B}{\partial \hat{x}\partial \hat{y}^2}\right)\\
 &&+\frac{D_2 k^2}{2 \pi} \left(-\frac{\partial^2 A}{\partial \hat{y}^2} +\frac{\partial^2 B}{\partial \hat{y}^2} - 2\hat{y} \frac{\partial^3 B}{\partial \hat{y}^3} \right), 
 \notag
\end{eqnarray}
with $u_3=0$,  where the functions $A$ and $ B$ are given by
\begin{align}
 A & = \frac{1}{2} \ln{[\cosh{(\hat{y}-\hat{y}_0)}-\cos{(\hat{x}-\hat{x}_0)}]},\\
 B &=\frac{1}{2} \ln{[\cosh{(\hat{y}+\hat{y}_0)}-\cos{(\hat{x}-\hat{x}_0)}]}.
\end{align}
This is an exact analytic solution for a periodic array of source dipoles near a no-slip wall. The corresponding  streamlines are shown in  Fig.~\ref{fig:4}. The net two-dimensional flow rate pumped along the swarm edge, $q$, can be evaluated analytically and we obtain 
\begin{align}
q&=\int_0^{+\infty}u_1 dy =\frac{D_1 k}{2 \pi}=\frac{D_1}{L},
\end{align}
with a total  flux in the thin fluid layer, $Q$,  given by 
\begin{align}
Q&=\int_0^{H}q dz=\int_0^{H} \frac{D_1 k}{2\pi} dz=\frac{Hk}{2\pi} \int_0^{1} D_1(\hat{z}) d\hat{z},
\end{align}
which can be evaluated and leads to
\begin{align}
Q&=\frac{Hk}{4\pi\mu}\left[F_1 H \hat{h}(1-\hat{h})+G_2\left(\frac{1}{2}-\hat{h}\right)\right].
\end{align}
Importantly,  we see in these expressions that only the $D_1$ component of the  source dipole (i.e.~the component parallel to the swarm edge at $y=0$) contributes to a net flow (a result which can also be seen by geometrical symmetry).  This means that the source dipoles due to a stokeslet and/or  a rotlet will create a net flux along the edge of the swam only due to their projection along the boundary at $y=0$. The orientation of the flagellar bundle will thus play a critical role in the creation of this flow. 

\subsection{Strength of flow singularities}

Knowing the fundamental solutions, and that they are source dipoles in the far field,  we now need to specify their strengths and spatial distribution  in order to get the complete picture on the flow. This requires first knowing the magnitudes of force and torque exerted by the flagellar bundle on fluid. 

\subsubsection{Resistive-force theory}
The force and torques can be estimated  using  resistive-force
theory (RFT) \cite{lighthill1976flagellar}. We denote by 
$\xi_{\perp}$ the drag coefficient for motion of the straight short filament perpendicular to its tangent and by
 $\xi_{\parallel}$  the drag coefficient in the parallel direction, and we write their ratio as  $\rho=\xi_{\parallel}/\xi_{\perp} < 1$. An elementary integration of the constant hydrodynamic force and moment densities along the helix leads to expressions for the  total force, $F$, and torque, $G$,  generated by the rotation of the flagellar bundle along the helical axis as
\begin{align}
\label{20} F&=\xi_{\perp} (a \omega - u_{fluid}) \Lambda (1-\rho) \sin{\Psi} \cos{\Psi},\\
\label{21} G&=\xi_{\perp} a (a \omega - u_{fluid}) \Lambda (\cos^2{\Psi}+\rho \sin^2{\Psi)},
\end{align}
where $a$ is the helix radius, $\Psi$ is the helix angle, $\omega$ is the angular speed of the rotating helix and $\Lambda$ is the arc length of the helix. In Eqs.~\eqref{20}-\eqref{21}, $u_{fluid}$ represents the magnitude of the background fluid velocity, which in fact may be safely neglected. Indeed, the typical local velocity of the rotating bundle of flagella is on the order of $a \omega\approx 300~\mu$m/s while the  fluid flows with typical speed 
$ u_{fluid}=10~\mu$m/s, and thus $ u_{fluid}\ll a \omega$.

\subsubsection{Physical parameters for \textit{E.~coli} bacteria}
The bacterial strains used in the  experiments from Ref.~\cite{wu2011microbubbles} are {\it E.~coli} HCB116; FliC S353C, antibiotics and arabinose were added to the swarm agar, $0.6\%$ Eiken agar, and experiments conducted at $30^{\circ}$C. Most of the  physical parameters  have been experimentally measured for {\it E.~coli} bacteria:
the pitch of the normal  LH helical bundle, $P=2.3~\mu$m \cite{turner2000real};  
the radius of the normal LH helical bundle, $a=0.3~\mu$m \cite{turner2000real};  
the axial length of the filament, $L_f=4.5~\mu$m$~\pm~2.0~\mu$m \cite{turner2010visualization}; 
the number of filaments per cell, $7.6 \pm 3.0$ \cite{turner2010visualization}; 
the radius of the bundle, $a_f=30$~nm \cite{turner2000real}; 
the mean cell length, $L_{cell}=5.2~\mu$m$~\pm~ 2.2~\mu$m \cite{darnton2010dynamics}; 
the cell diameter, $D_{cell}=1~\mu$m \cite{turner2010visualization}; 
the thickness of the fluid film $H=1~\mu$m \cite{wu2011microbubbles}. 

In addition to these geometrical parameters, the frequency of the flagellar bundle rotation is an important quantity. The chiral flow experiments were conducted at $30^{\circ}$C, but the flagellar rotational frequency was not measured. The flagellar rotation  frequency will depend strongly on the temperature in the experiment \cite{lowe1987rapid} as well as on the load applied on the flagella (flagella rotate in a thin film near boundaries) \cite{yuan2010asymmetry}. It has been reported  that the mean swarming and swimming speeds are about the same \cite{turner2010visualization}. More recently it was discovered that  bacteria increase the number of force-generating units that drive motors at high loads, suggesting that the rotational frequency of flagella in the swarm might be close to the frequency of the bacteria swimming in broth \cite{lele2013dynamics}. The estimated bundle frequency for cells at $32 ^{\circ}C$ is $f=156$~Hz \cite{lowe1987rapid} and given that the chiral flow experiments are carried out at $30 ^{\circ}C$ we estimate that the resulting  flagellar frequency is around $f=150$~Hz. In the remainder of the paper we thus take $f=150$~Hz as our model value, remembering that everything scales linearly with frequency.

\subsubsection{Drag coefficients}

We need to be careful in estimating the drag coefficients $\xi_{\perp}$, $\xi_{\parallel}$ because the flagellar filaments are located  near rigid boundaries. The values of the  drag coefficients were obtained numerically  by  Ramia \textit{et al}.~\cite{ramia1993role} using a general boundary-element method for the motion of a rod midway between two plates (parallel orientation).
We use these numerical  results to adjust the drag coefficient in the infinite fluid, denoted $\xi_{\perp \infty}$ and $\xi_{\parallel \infty}$.
If we take the thickness of fluid film to be  $H=1~\mu$m and the length of the flagella bundle $L_f=4.5 \mu m$ then we have 
\begin{align}
 \xi_{\parallel} &=\gamma \xi_{\parallel \infty}=1.7 \xi_{\parallel \infty},\\
 \xi_{\perp} &=\gamma_1 \xi_{\parallel}=\gamma_1\gamma \xi_{\parallel \infty}=4.25\xi_{\parallel \infty},\\
 \rho &=\frac{ \xi_{\parallel}}{ \xi_{\perp} }=\frac{1}{\gamma_1}=0.4,
\end{align}
where $\gamma = 1.7$ and $\gamma_1=2.5$ are taken from Ramia \textit{et al}.~\cite{ramia1993role}. To estimate the values of the  drag coefficients in the infinite fluid, we use the results obtained by Lighthill \cite{lighthill1976flagellar} 
\begin{align}
 \frac{\xi_{\parallel \infty}}{\mu} &=\frac{2\pi}{\ln{(0.18 P/a_f)}}=2.39,
\end{align}
which, combined with Ramia's results, leads to
\begin{align}
  \frac{\xi_{\parallel}}{\mu}&=1.7\times 2.39=4.07,\\
  \frac{\xi_{\perp}}{\mu} &=4.25\times 2.39=10.17,\\
 \rho &=0.4.
\end{align}
These values capture the fact that it is harder  to move near no-slip walls than  moving in the infinite fluid, and that it is even harder to move
perpendicularly than to move along walls giving larger values for $\xi_{\perp}$ and a smaller value for  $\rho$.

\subsection{Distribution of singularities}
There are different ways to model the flagellar bundle and its action on the fluid, from simple to more complex. We consider here four different levels of modeling for the flow field due to the rotating flagellar bundle between two no-slip walls.  Firstly, we model the helical bundle as single point force and torque located at the center of the bundle axis, giving the time-averaged flow field. A second approach models  the helix as a line distribution of forces and 
torques along the helical axis, all with equal strengths and thus giving also a time-averaged flow field. Third, we can use a helical distribution of
equal-strength  point forces distributed along the centerline of the flagellar bundle to capture the instantaneous flow field, which can again be averaged over time. Finally, we can derive an approximate analytical result for this helical distribution in the asymptotic  limit when the helix radius is small compared with its pitch.

\subsubsection{A single point force and torque}
\begin{figure}[t]
\flushleft
\hspace*{.8in}
\includegraphics[width=0.65\textwidth]{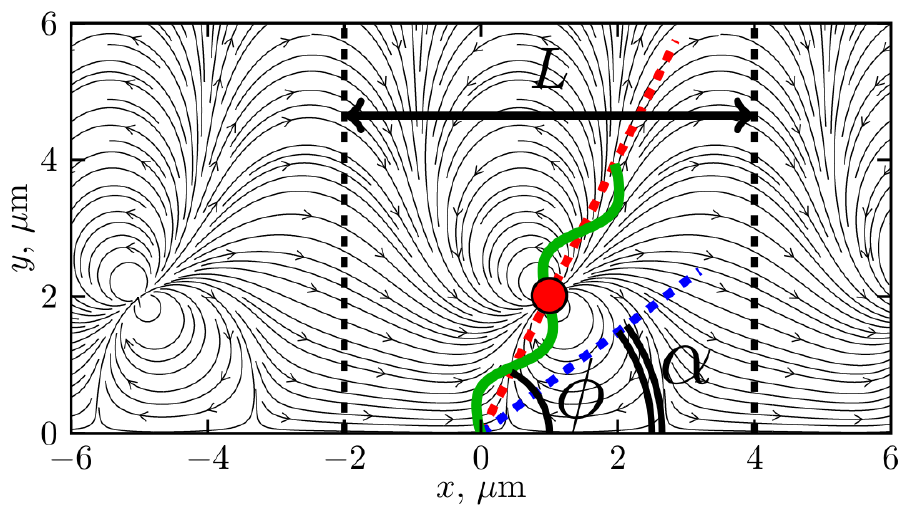}
\hspace*{.8in}
\includegraphics[width=0.78\textwidth]{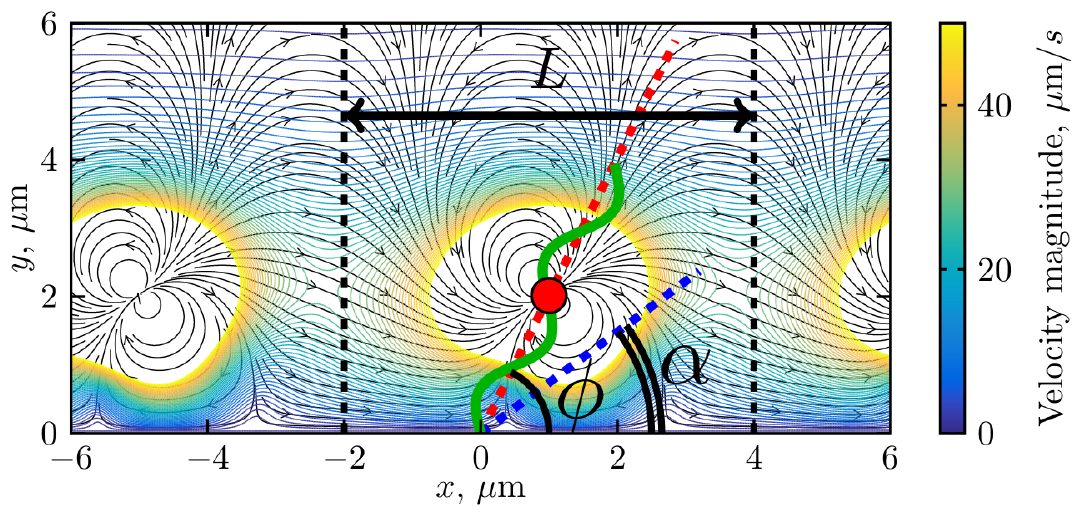}
\caption {Streamlines and velocity contour lines for the flow field due to a single two-dimensional source dipole (red dot) with periodic boundary conditions and no-slip at the swarm edge ($y=0$). The red dotted line shows the direction of the bundle of flagellar filaments and the blue dotted line shows the direction of the resulting source dipole (combination of force and torque). The black dashed vertical lines show the midpoint between two singularities.  The axial length of the flagellar bundle is $L_f=4.5$~$\mu$m, the period length is $L=6$~$\mu$m, the scaled height of the flagella bundle axis above the agar is $\hat{h}=h/H=0.35$, the flagellar frequency is $f=150~$Hz, the bundle angle and the source dipole direction with the swarm edge are $\phi=63.6^\circ$, $\alpha=36.6^\circ$ respectively. The source dipole strengths are $D_x=505~(\mu$m$)^3/s$ and $D_y=375~(\mu$m$)^3/s$. The velocity magnitude contours are cut off at $50~\mu$m$/s$.} 
\label{fig:5}
\end{figure}

The simplest case would be to represent the flagellar bundle with a single point force, $\vec{F}$, and torque, $\vec{G}$, located at the centre of the bundle axis. The flow due to a point force and torque located between two parallel walls results in a source dipole at the leading order. 
In this case, we would then place a single source dipole at
\begin{align}
      x_0&=\frac{L_f}{2}\cos{\phi},\\
      y_0&=\frac{L_f}{2}\sin{\phi},\\
      z_0&=h,
\end{align}
where $L_f$ is the flagellar bundle length along its axis and $\phi$ is the angle between the flagellar axis and the $x$ axis.
The streamlines and velocity contour lines for this flow with periodic boundary conditions are shown in Fig.~\ref{fig:5} for $L_f=4.5$~$\mu$m,  $\phi=63.6^\circ$, $\hat{h}=h/H=0.35$, $f=150$~Hz and a period of $L=6$~$\mu$m. The resulting source dipole strengths are $D_x=505~(\mu$m$)^3/s$ and $D_y=375~(\mu$m$)^3/s$.

\subsubsection{A line distribution of forces and torques}
A more realistic model would have a uniform line distribution of forces with constant force density $\vec{F}/L_f$ and torques with constant torque density $\vec{G}/L_f$ along the axis of the helical bundle of filaments.
This would then lead to a uniform line distribution of source dipoles along the flagellar bundle axis at locations
\begin{align}
      x_0&=P l\cos{\phi},\\
      y_0&=P l\sin{\phi},\\
      z_0&=h,
\end{align}
where  the parameter $l$ varies  from $l=0$ to $l=L_f/P$.

\subsubsection{A helical distribution of point forces}
\begin{figure}[t]
\centering
\includegraphics[width=0.78\textwidth]{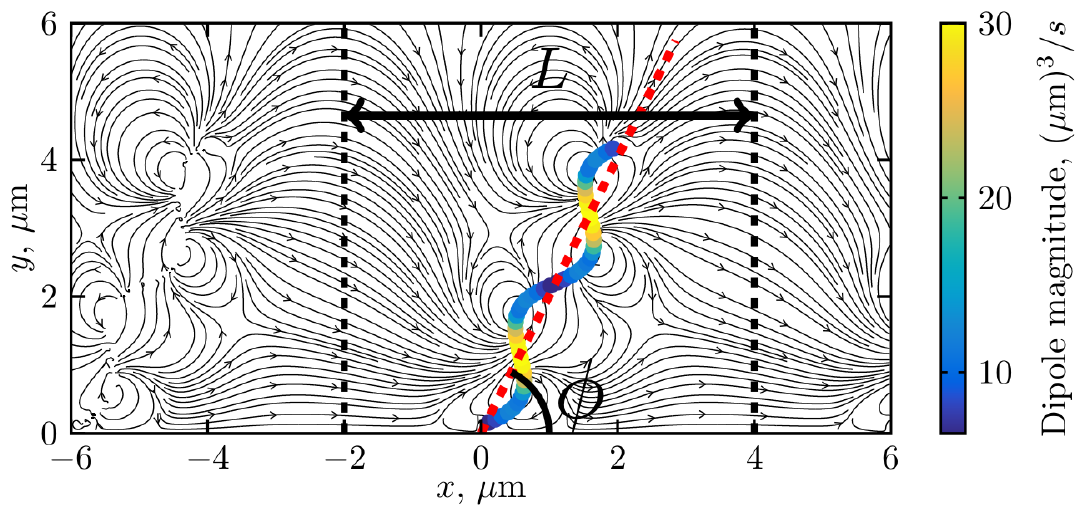}
\includegraphics[width=0.78\textwidth]{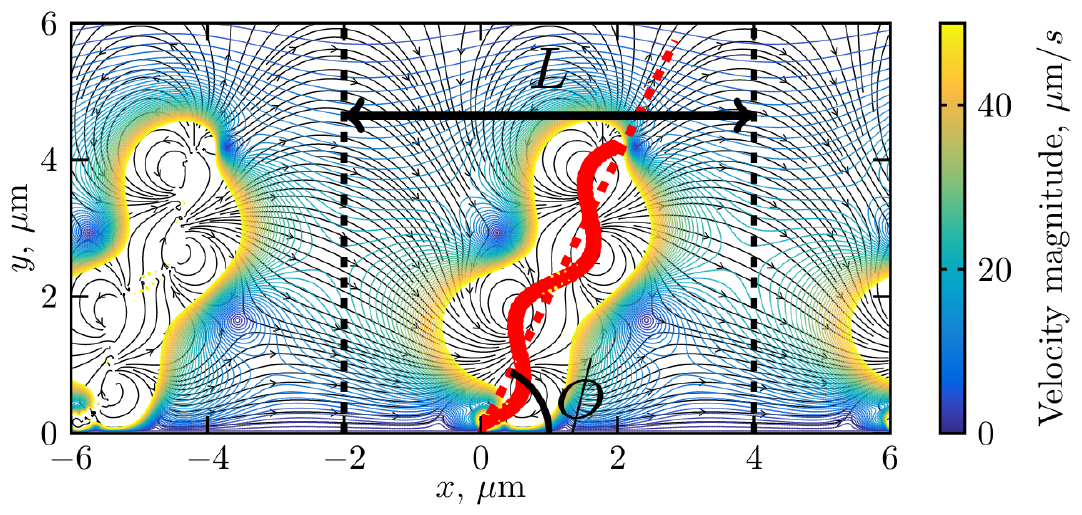}
\caption {Left: streamlines of the instantaneous flow field due to the helical distribution of point forces along the flagellar bundle. The coloured dots represent the two-dimensional source dipole locations and strengths. 
Right: contour lines of the instantaneous velocity magnitude due to the helical distribution of point forces along the flagellar bundle.
The parameters are the same as in Fig.~\ref{fig:5}, namely $L_f=4.5$~$\mu$m, $L=6$~$\mu$m, $\hat{h}=h/H=0.35$, $f=150~$Hz and $\phi=63.6^\circ$. The pitch of the normal LH helical bundle is $P=2.3~\mu$m and the radius of the helical bundle is $a=0.3~\mu$m. The number of source dipoles along the bundle is $N=51$. The velocity magnitude contours are cut off at $50~\mu$m$/s$.}  
\label{fig:6}
\end{figure}

\begin{figure}[t]
\centering
\includegraphics[width=0.78\textwidth]{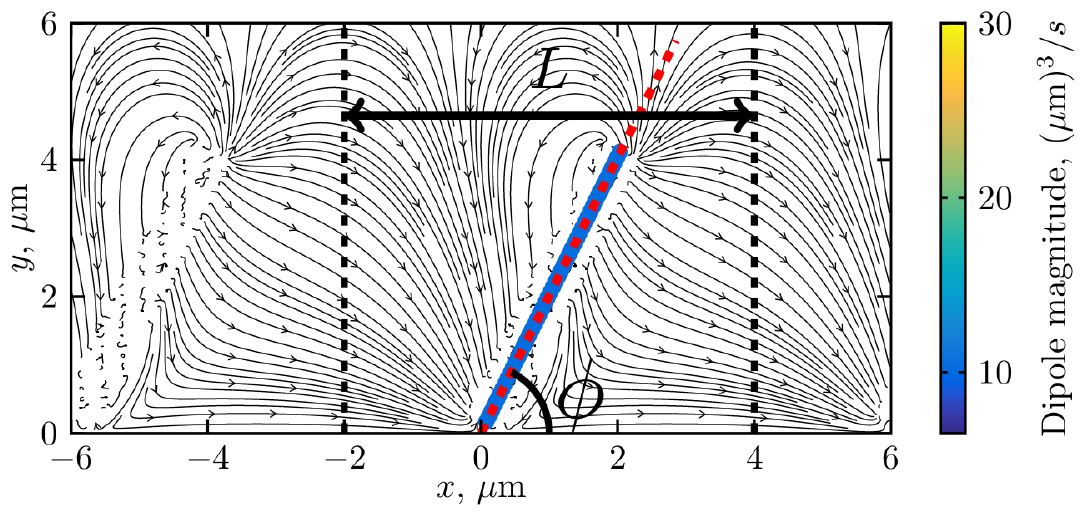}
\includegraphics[width=0.78\textwidth]{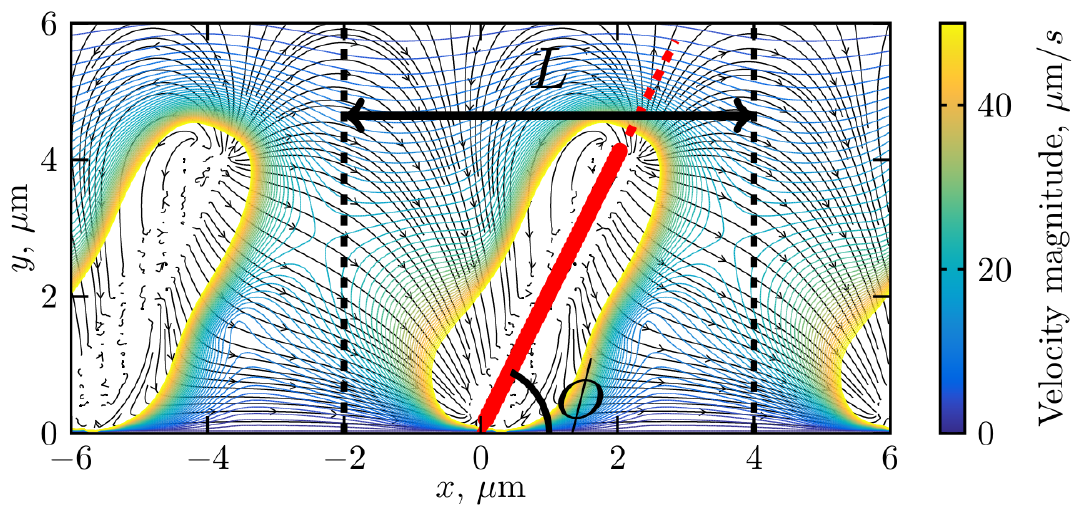}
\caption {Same as in Fig.~\ref{fig:6} but after time-averaging over one period of rotation of the flagella.}
\label{fig:7}
\end{figure}

To fully capture   the geometry of the helix, we can instead elect to consider a helical distribution of flow singularities. The 
centerline of the left-handed helix with  radius $a$ and  pitch $P$ can be described  as
\begin{align}
      x_0(t)&=P l\cos{\phi} + a \sin{(\omega t - 2\pi l)}\sin{\phi},\\
      y_0(t)&=P l\sin{\phi}-a \sin{(\omega t - 2\pi l)}\cos{\phi},\\
      z_0(t)&=h+a\cos{(\omega t - 2\pi l)},
\end{align}
where  $\omega$ is the constant angular speed and $t$ is time. The parameter $l$ varies  from $l=0$ to $l=L_f/P$.  The arc length of the small segment $\delta l$ of the helix can be written as
\begin{eqnarray}
  \delta s&=&\sqrt{(dx_0)^2+(dy_0)^2+(dz_0)^2}\\
  &=& \sqrt{P^2+4\pi^2 a^2} \delta l. \notag
\end{eqnarray}
Along the small segment $\delta s$ one can apply RFT to calculate the  force density acting on the fluid,  $\delta\vec{F}$, as 
\begin{equation}
  \delta \vec{F}=\left(\xi_{\parallel} \vec{t} \vec{t} + \xi_{\perp}\vec{n} \vec{n}\right)\cdot \vec{u}_{S} \delta s,
\end{equation}
where $\vec{t}$ is the unit tangential vector along the segment, $\vec{n}$ is the unit normal to the segment and $\vec{u}_S$ is the velocity of the flagellar centerline. Note that $\hat{z}_0=z_0/H=\hat{z}_0(t)$ is a function of time $t$ and thus  the strengths of the source dipoles along and across the helix axis will change as the helix is rotating, namely
\begin{align}
  \delta D_{\parallel}(t) &\propto  \hat{z}_0(t) \left(1-\hat{z}_0(t)\right),\\
  \delta D_{\perp}(t) &\propto \cos{(\omega t - 2\pi l)} \cdot \hat{z}_0(t) \left(1-\hat{z}_0(t)\right).
\end{align}
This helical distribution of point forces in three dimensions, or the sinusoidal distribution of source dipoles in two dimensions, 
gives the instantaneous flow field $\vec{u}(t)$. The time-averaged flow field is found by averaging over one period $T=2\pi/\omega$. The contour lines for the instantaneous flow field are shown in Fig.~\ref{fig:6} and for the time-averaged flow field in Fig.~\ref{fig:7}.

\subsubsection{Modified line distribution}
The time averaged flow field can be found approximately analytically in the following limit. Let us consider the case where the helix radius is much smaller than its pitch, i.e.~$a/P \ll 1$ (for \textit{E.coli} bacteria $a/P=0.13$). In that limit, we get at leading order in $a/P$ the helix centerline at
\begin{align}
      x_0& \approx P l\cos{\phi},\\
      y_0& \approx P l\sin{\phi},\\
      z_0(t)&=h+a\cos{(\omega t - 2\pi l)}.
\end{align}
We can next average analytically in that geometry over time $t$ to obtain the time-averaged flow field. It is the flow due to the line distribution of source dipoles with strengths
\begin{align}
 D_{\parallel}(\hat{z})&=\frac{3}{\mu} F_{\parallel} H \left[\hat{h}\left(1-\hat{h}\right)-\frac{a^2}{2H^2}\right] \hat{z}  \left(1-\hat{z}\right),\\
 D_{\perp}(\hat{z}) &=\frac{3}{\mu} F_{\perp}a \left(\frac{1}{2}-\hat{h}\right) \hat{z}  \left(1-\hat{z}\right),
\end{align}
where
\begin{align}
 F_{\parallel}&=\xi_{\perp} a \omega \Lambda (1-\rho) \sin{\Psi} \cos{\Psi},\\
 F_{\perp}&=\xi_{\perp} a \omega \Lambda (\cos^2{\Psi}+\rho \sin^2{\Psi)}.
\end{align}
In this approximate model, the source dipole in the direction perpendicular to the helix axis is the same as the one derived using RFT with the point torque. As a difference, the component of the  source dipole parallel to the helix axis is modified by a factor $a^2/2H^2$ which captures the finite size  of the helix compared to the film thickness. 

\subsection{Measuring the tangential flow}
In order to eventually compare the time-averaged  tangential flow $u(x,y,z)$ with experimental results we need to choose where we  measure the flow, i.e.~choose values for $x$ and $ z$. In the experiments, the tangential flow is given by a bubble moving near the swarm edge which could be anywhere across the thin film. We will thus assume to a leading order that this bubble translates as a passive
small tracer with the mean velocity. In our model, the flow is parabolic in $\hat z$, with a dependence of $\hat{z}(1-\hat{z})$, and therefore averaging across the flow we get
\begin{equation}
  \int_0^1 \hat{z}(1-\hat{z}) d\hat{z}=1/6.
\end{equation}
We will denote the  velocity profile resulting from the $z$ and time-averaging  as $\langle u \rangle (x,y)$. Furthermore, by mass conservation, the mean flow rate in the $y,z$  plane is the same for every value of $x$, and for convenience we  choose $x$ to be in the middle between two flagellar bundle centres, as indicated by dashed lines in Figs.~\ref{fig:5}-\ref{fig:7}.  The final tangential flow speed of interest will then be 
denoted $\langle u \rangle (y)$.  

\subsection{Comparing the models}

\begin{figure}[t]
\centering
\includegraphics[width=0.6\textwidth]{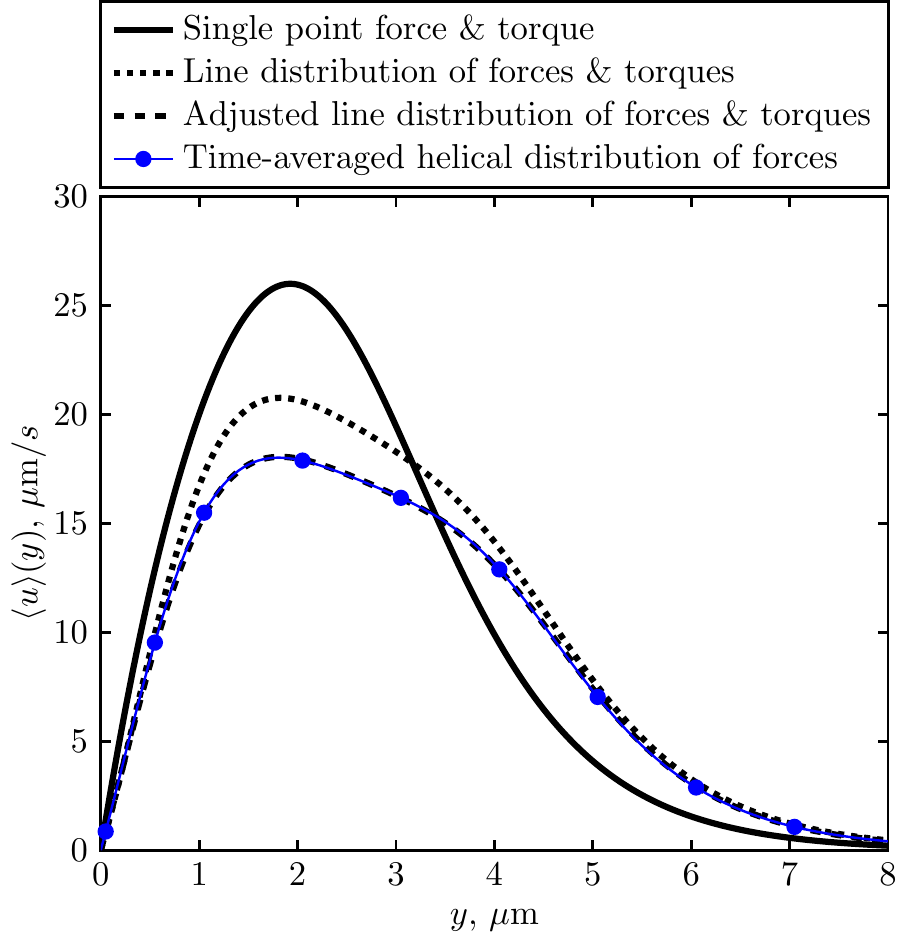}
\caption {Tangential speed profiles, $u(y)$, using four different singularity distributions:
a single point force and torque; a line  distribution of forces and  torques; the modified   line distribution of forces and torques; and the time-averaged helical distribution of forces.  The flow predicted by the  adjusted line distribution is shown to perfectly match with the one produced by the averaged helical distribution. The parameters are: 
$\phi=67.25^{\circ}$, $L=8~\mu $m, $\hat{h}=0.35$, $f=150$~Hz.} 
\label{fig:8}
\end{figure}

The effectiveness of the four models is displayed in 
 Fig.~\ref{fig:8} where we plot the mean tangential velocity, $\langle u\rangle$ (in $\mu$m/s) as a function of the distance, $y$, away from the swarm edge for the four models (single singularity; line distribution;  helical distribution and modified line distribution). Importantly, we see the excellent quantitative agreement between the time-averaged helical distribution and the steady, modified line-distribution of source dipoles.

%%%%%%%%%%
\section{Results and comparison with experiments}
\subsection{Qualitative understanding of the flow}
In order to gain intuition about the main features of the flow created by the flagella, we can use the simplest model of a periodic array of point forces and torques. This has the advantage of giving a simple analytical expression for the averaged flow along the swarm.
If we write $k(x-x_0)=\pi$ (i.e.~we measure the flow in the middle between two bundles) then we obtain the tangential flow speed, $\langle u\rangle (y)$, as
\begin{equation}
 \langle u\rangle (y)=\frac{k^2}{4 \pi} (D_F\cos{\phi}+D_G\sin{\phi}){\Delta} ,
\end{equation}
where ${\Delta}$, $D_F$ and $D_G$ are given by
\begin{eqnarray}
\Delta& =  & \frac{1}{1+\cosh{(\hat{y}-\hat{y}_0)}}  -\frac{1}{1+\cosh{(\hat{y}+\hat{y}_0)}} \notag \\
&& + \frac{2\hat{y} \sinh{(\hat{y}+\hat{y}_0)}} {[1+\cosh{(\hat{y}+\hat{y}_0)}]^2}, \\
 D_F &= &\frac{1}{2\mu}F H \hat{h}\left(1-\hat{h}\right),\\
 D_G &=& \frac{1}{2\mu}G \left(\frac{1}{2}-\hat{h}\right).
\end{eqnarray}

\begin{figure}
        \includegraphics[width=0.47\textwidth]{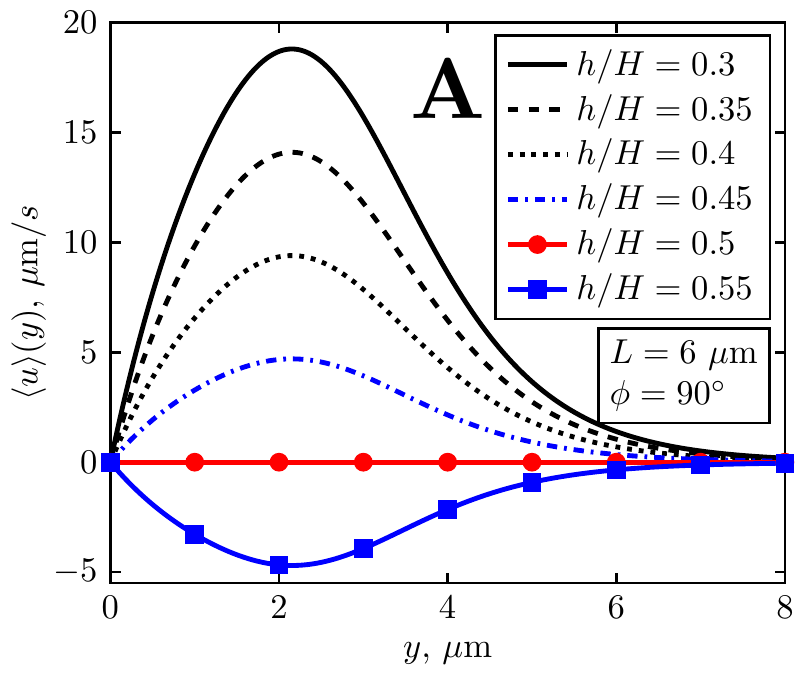}
        \includegraphics[width=0.47\textwidth]{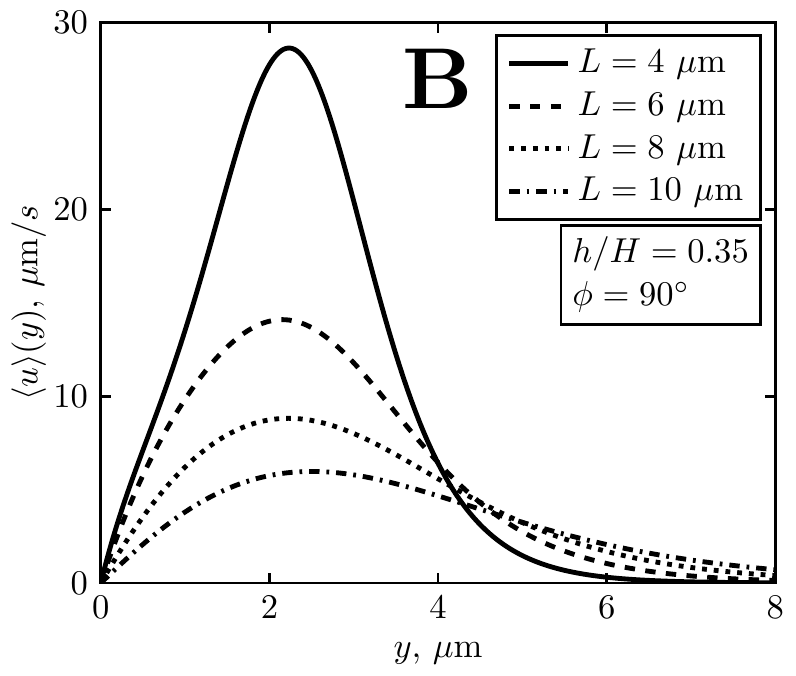}
        \includegraphics[width=0.47\textwidth]{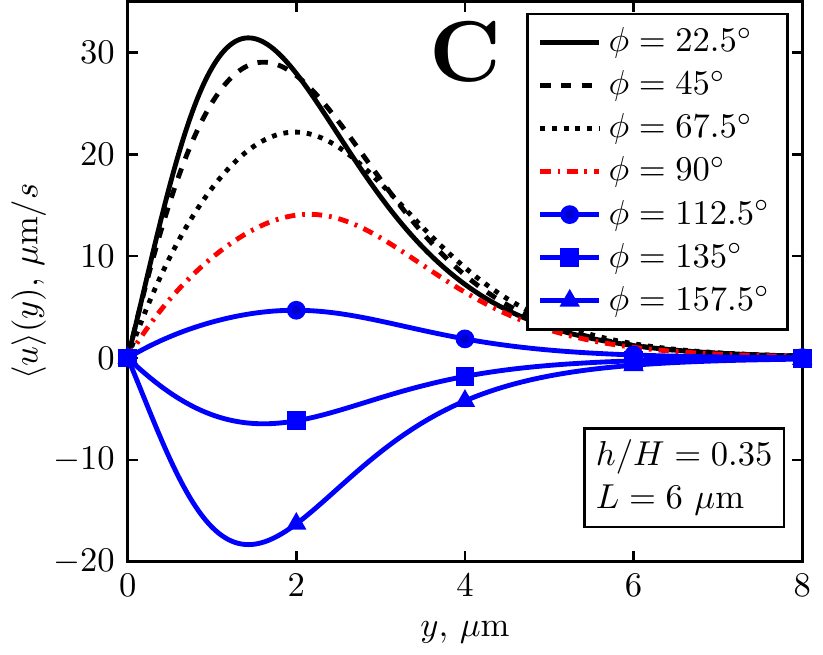}
        \includegraphics[width=0.47\textwidth]{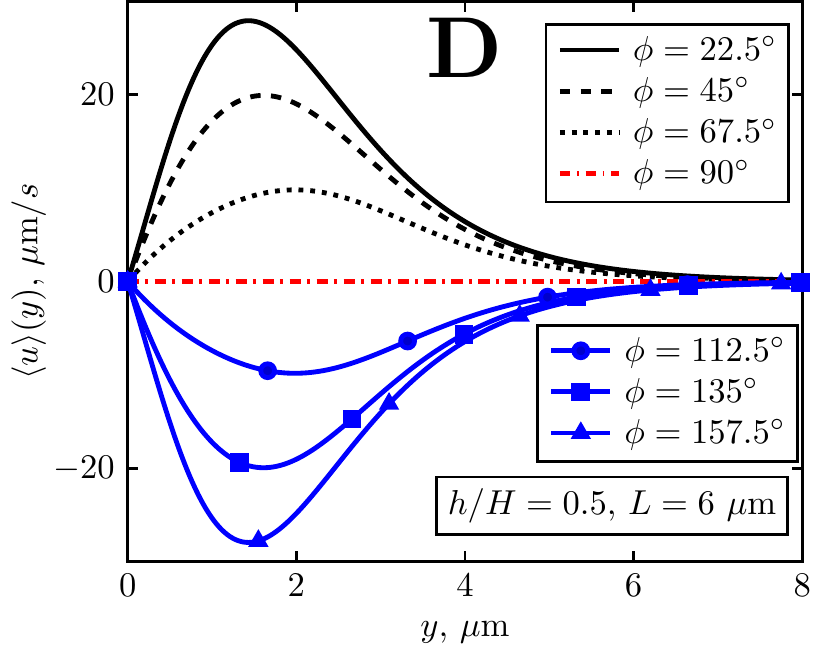}
       \caption {Velocity profiles along the swarm edge, $\langle u\rangle (y)$,  for a rotation frequency of the flagella bundle of $f=150$~Hz. \textbf{A}: Dependence on the bundle height $\hat{h}=h/H$ with $\phi=90^{\circ}$ and $L=6~\mu$m; \textbf{B}:  Dependence on the period $L=2\pi/k$ with $\phi=90^{\circ}$ and $\hat{h}=0.35$;       \textbf{C}: Dependence on the bundle angle $\phi$, with $\hat{h}=0.35$ and $L=6~\mu$m; \textbf{D}: Dependence on the bundle angle $\phi$ with  $\hat{h}=0.5$ and  $L=6~\mu$m.} 
      \label{fig:9}
\end{figure}

\begin{figure}[t]
\centering
\includegraphics[width=0.6\textwidth]{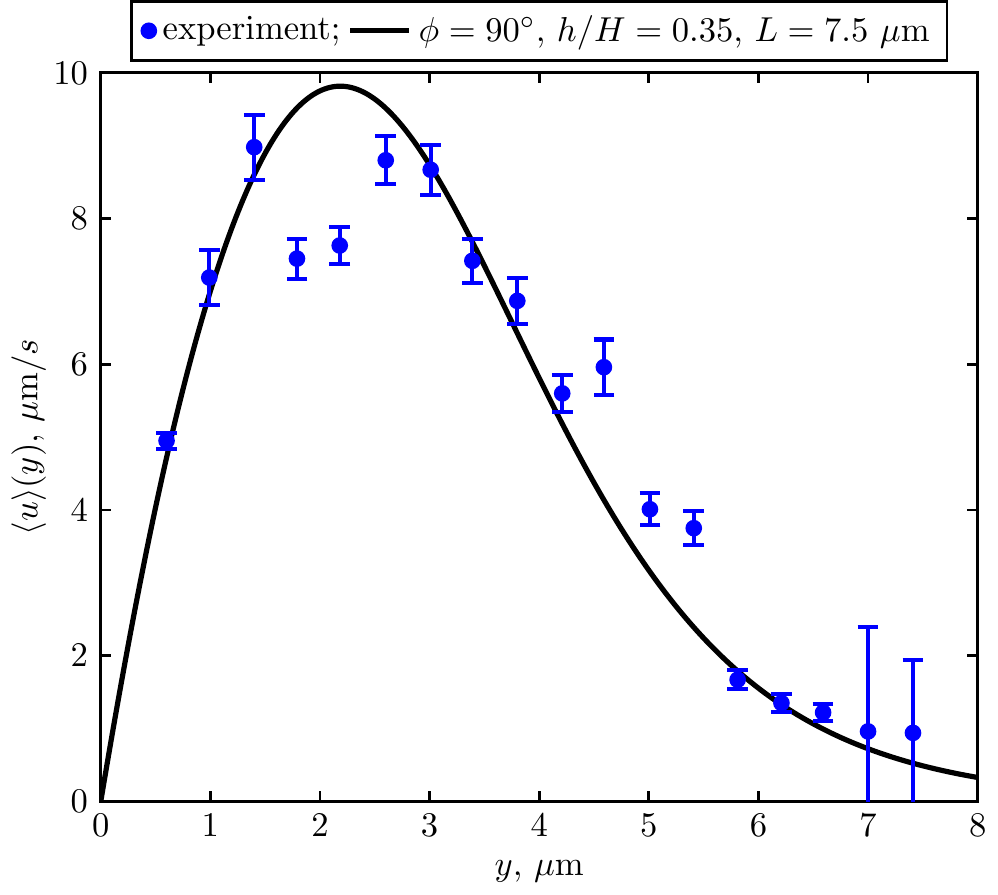}
\caption {Tangential velocity profile using the model with the single point force and torque for one particular set of parameters: flagellar rotation frequency $f=150$~Hz;  angle $\phi=90^{\circ}$; 
 height $\hat{h}=h/H=0.35$;  period length $L=7.5~\mu$m. The resulting vertically averaged source dipole strength parallel to the boundary is $D_1\approx 290~(\mu$m$)^3/s$, and perpendicular to the boundary is $D_2\approx 560~(\mu$m$)^3/s$. The flux generated around the swarm is $Q=H  q=H  D_1/L\approx 38~(\mu$m$)^3/s$.}
\label{fig:10}
\end{figure}

The resulting  velocity profiles along the swarm edge, $\langle u\rangle (y)$, are plotted in Fig.~\ref{fig:9}, for a variety of   different parameters, showing the dependence on the vertical position of the flagella 
(Fig.~\ref{fig:9}A), the length between each flagellar bundle (Fig.~\ref{fig:9}B), the angle between the flagella and the swarm edge (Fig.~\ref{fig:9}C and D). We also illustrate one particular fit of the single-singularity model to the experimental data  in Fig.~\ref{fig:10}; a more systematic fitting approach will be carried out in the next section to gain further biological insight.

Since the flow has the dependence $D_F \sim \hat{h}(1-\hat{h})$,  the maximum flow induced by the force is attained at $\hat{h}=1/2$, i.e.~when the axis of the flagellar bundle 
is in the mid-plane  between the  two interfaces. By contrast,  $D_G \sim (1/2-\hat{h})$,  which leads to larger flow values when the flagellar  axis is closer to one of the interfaces,  and is  exactly zero if there is a symmetric situation with $\hat{h}=1/2$ (see Figs.~\ref{fig:9}A and D.) 

As shown in Fig.~\ref{fig:9}B, the  period $L=2\pi/k$, modeling the typical  distance between flagella on different cells, sets the width of the tangential speed profile as well as  the maximum amplitude of the velocity profile.

The  angle $\phi$ between the helical axis and the swarm edge  modifies (a) the relative importance between the force and the torque contribution to the net flow  and (b) the location of the point force, $\hat{y}_0=L_f \sin{\phi}/2$, as illustrated in  Fig.~\ref{fig:9}C. The results in Fig.~\ref{fig:9}D  further show the situation when the  point force and torque are exactly in the mid-plane between the two interfaces, 
i.e.~$\hat{h}=0.5$. By symmetry, the torque does not contribute to any flow and a perfect antisymmetry in the flow is obtained between situations where $\phi<90^\circ$ and $\phi > 90^\circ$.

To understand  qualitatively the relative importance of the force vs.~the torque in the produced flow, we can use the analytical formulas derived above. First, using  RFT and the 
the experimental values for \textit{E.~coli} stated earlier in the paper, we can  compute the approximate magnitude of the flagellar force, $F$, times the film
thickness $H$ and that  of the torque $G$. We obtain
\begin{align}
 F  H&=\xi_{\perp} H a \omega \Lambda (1-\rho) \sin{\Psi} \cos{\Psi} \nonumber\\
 &\approx 4.92~ \text{pN}\mu\text{m},
 \end{align}
and
\begin{align}
 G&=\xi_{\perp} a^2 \omega \Lambda (\cos^2{\Psi}+\rho \sin^2{\Psi)} \nonumber\\
 &\approx 3.81~\text{pN}\mu\text{m},
\end{align}
so their ratio is
\begin{equation}\label{56}
 \frac{F H}{G}\approx 1.3.
\end{equation}
In addition to this  $30\%$ difference between the 
 contribution due to the force and that due to the  torque,  the  strengths of the source dipoles depend also on the value of $\hat{h}$ and flagellar angle $\phi$. The ratio between source dipole strengths due to force and due to torque contributing to the tangential flow is
\begin{equation}
R_{FG}=\frac{\abs{D_F\cos{\phi}}}{\abs{D_G \sin{\phi}}}=\frac{FH\hat{h}(1-\hat{h})}{G\abs{1/2-\hat{h}}}\abs{\cot{\phi}},
\end{equation}
and given Eq.~\eqref{56} this leads to
\begin{equation}
R_{FG}\approx 1.3\frac{\hat{h}(1-\hat{h})}{\abs{1/2-\hat{h}}}\abs{\cot{\phi}}.
\end{equation}
If $R_{FG} \ll 1$ then the tangential flow is mainly generated by the torque applied on the fluid. If $R_{FG} \gg 1$ then the tangential flow results from the force applied on the fluid. Notably, if the flagellar bundle is pointing radially out of the swarm, $\phi=90^\circ$, then $R_{FG}=0$ while if $\phi=0^\circ$ or $180^\circ$ then $R_{FG}^{-1}=0$.

In our model with two no-slip surfaces, the flagellar bundles have to be closer to the agar surface than the fluid/air interface in order to generate the chiral flow in the clockwise direction because the flow due to a point torque is proportional to $(1/2-\hat{h})$, i.e.~asymmetric around $\hat{h}=1/2$. If the top fluid/air interface  was not immobile but instead  satisfied a no-shear boundary condition  then the flow due to the torque applied on the fluid would be proportional to $(1-\hat{h})$ and it would not change sign anywhere through the fluid ($0<\hat{h}<1$, see Appendix \ref{A}). This means that in this situation the chiral flow would always be generated in the clockwise direction due to the torque applied. 

Building on the good agreement between the simple point-singularity model and the experimental data in Fig.~\ref{fig:10}, we proceed in the next section to use the  more realistic line model in order to gain further quantitative insight into the biological system.

\begin{figure*}[t]
        \includegraphics[width=0.325\textwidth]{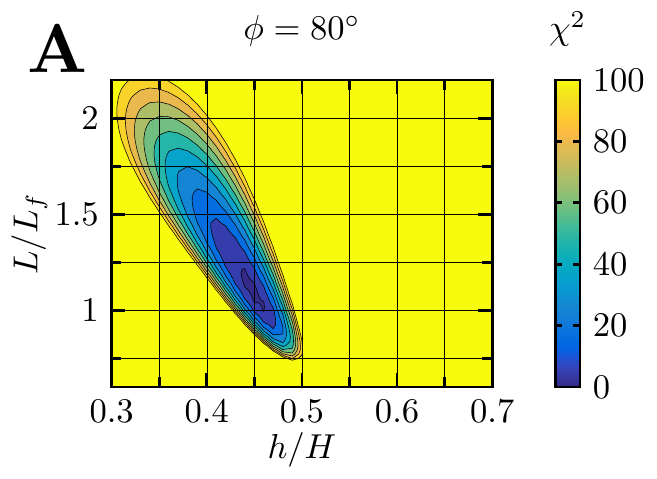}
        \includegraphics[width=0.325\textwidth]{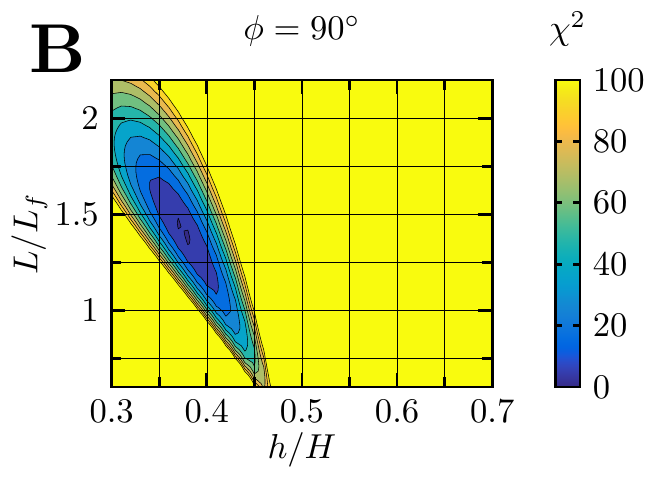}
        \includegraphics[width=0.325\textwidth]{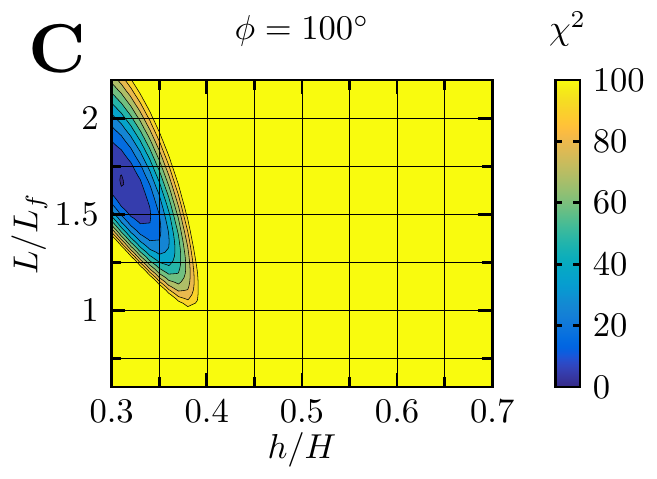}
        \includegraphics[width=0.53\textwidth]{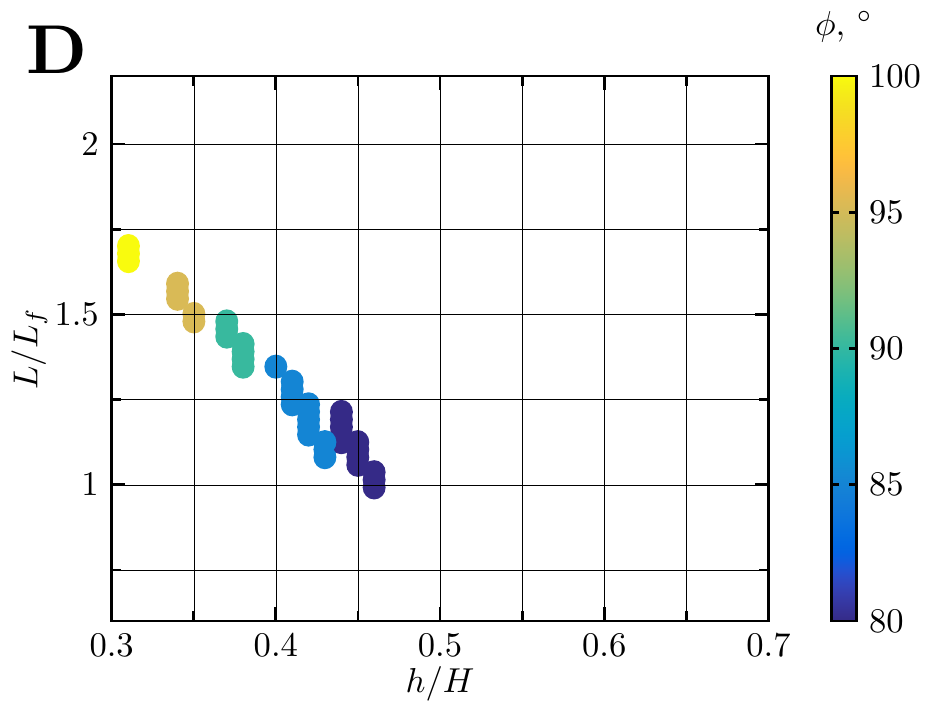}
        \includegraphics[width=0.46\textwidth]{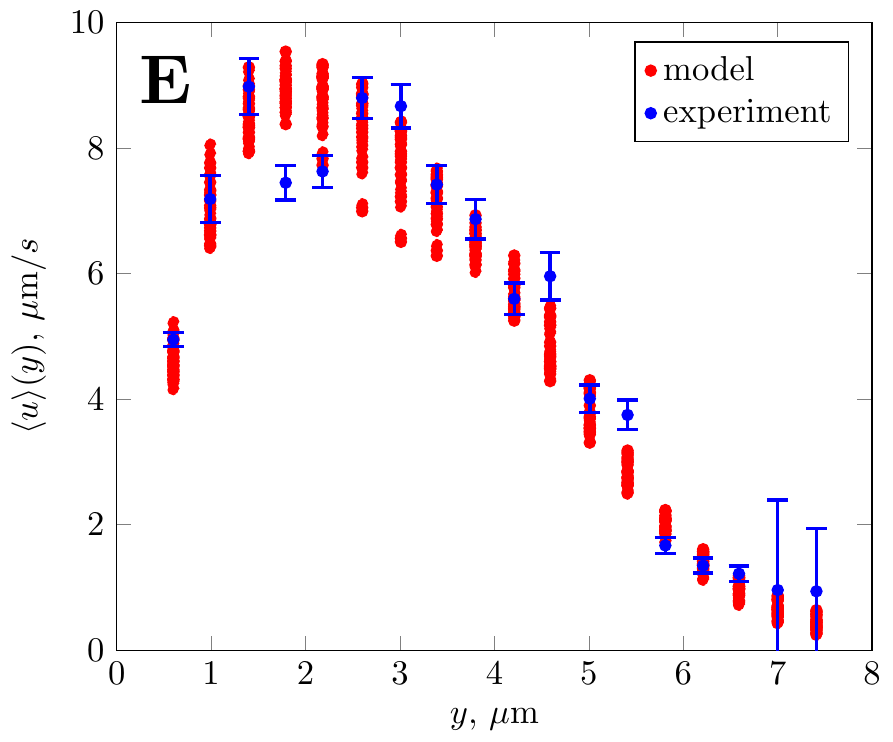}
        \caption {Quantitative three-parameter fit of our modified line distribution model to the experimental data of Ref.~\cite{wu2011microbubbles}. The three fitting parameters are:  the height $\hat{h}$, the period $L$ and the angle $\phi$. The other (fixed) parameters are the  frequency $f=150$~Hz,  helix length $L_f=4.5~\mu$m, and  fixed fluid film thickness $H=1~\mu$m. The figures \textbf{A}, \textbf{B}, \textbf{C} show contour plots of $\chi^2$ for $\phi=80^{\circ}$, $\phi=90^{\circ}$, $\phi=100^{\circ}$ respectively (no fit with $\chi^2 < 10$ was obtained outside this range of angles). The cut-off is at $\chi^2=100$ (yellow area). The figure \textbf{D} shows the scatter plot of parameter combinations $L/L_f$, $h/H$, $\phi$ such that $\chi^2 <10$. The best fit has $\chi_{min}^2=7.41$. The figure \textbf{E} shows the tangential speed profiles due to the flows with parameter combinations such that $\chi^2 <10$ superimposed with the experimental data.} 
        \label{fig:11}
\end{figure*}

\subsection{Quantitative fitting of experimental data}
The modified line-distribution model can be used to quantitatively fit the experimental data and infer some of the relevant  physical characteristics of  the biological system.  We use  chi-squared statistics to determine the goodness-of-fit of the data from Ref.~\cite{wu2011microbubbles} (reproduced in Fig.~\ref{fig:2}) to the model, and  want thus to minimize
\begin{align}\label{chi}
 \chi^2=\frac{1}{N-p}\sum_{i=1}^{i=N} \frac{(x_i-\mu_i)^2}{\sigma_i^2},
\end{align}
where the data with $N=18$ experimental points gives the mean values ($x_i$) and the standard deviations ($\sigma_i$) while  the model provides $\mu_i$.  In Eq.~\eqref{chi}, $p$ is the number of   free parameters in the model, which  is $p=3$:  the flagella angle $\phi$, the vertical position of the axis of flagella $\hat{h}=h/H$ and the period rescaled by the flagella length
$L/L_f$. We test the whole range of geometrically feasible parameters: $0.6<L/L_f<2.2$, $0.3<h/H<0.7$, $5^{\circ}<\phi<175^{\circ}$. All other  parameters are fixed:  flagellar frequency of $f=150$~Hz, the thickness of the fluid film  $H=1~\mu$m, and 
the flagella length $L_f=4.5~\mu$m. The flagellar bundles are represented by the modified line of source dipoles.  We then plot in Fig.~\ref{fig:11} the contour lines for the chi-squared, $\chi^2$, (A-C); a scatter plot of the flagellar angles for all best fits with $\chi^2<10$ (it is impossible to find fits with $\chi^2<10$ for flagella angles outside the range 80$^{\circ}$-100$^{\circ}$) (D); and  the corresponding predicted profile for the flow velocity along the swarm edge for each angle choice (E).

From the range of parameters able to  fit the data we learn an important biophysical fact, namely that  the flagellar filaments  are mostly pointing radially out of the swarm and therefore there  is no wrapping  of the flagella around the cells or near the edge. 
We further learn that the distance between  bundles  of flagella contributing to the flow, or the period, is in the range of $L\approx 5-8~\mu$m. This  corresponds to the situation where approximately one cell is aligned along the swarm edge between each bundle of flagella since the cell length has been measured experimentally to be  $L_{cell}=5.2\pm2.2~\mu$m \cite{darnton2010dynamics}, a result fully consistent with  recent experimental observations of cellular orientations near swarms  \cite{copeland2010studying}. Finally, we see that in order to produce this CW chiral flow we need to have the left-handed flagellar bundle just below the center line closer to the agar surface, i.e.~$\hat{h}={h}/{H}$ in the range $0.3-0.45$. 

These results have implications for biology. The biological hypothesis that the observed chiral flow is driven by the flagella sticking out of the swarm is fully consistent with the predictions of our model. We predict that the  flagella are not wrapped around the cells, in contrast to what was found for sparse distributions of bacteria above the agar-creating circulations around the individual cells \cite{wu2011microbubbles} and for stuck cells \cite{cisneros08}.  According to our results, the flagella are pointing almost radially out of the swarm contributing to the fluid movement in the radial direction due to pushing on the fluid and  creating the chiral flow in the clockwise direction due to the torque applied to the fluid. While fluid-structure interactions would encourage radially-oriented flagella to be bent and swept tangentially by the chiral flows, our results indicate that the flow  speeds might not be large enough to lead to significant flagellar bending.

The direction of the chiral flow is clockwise because the left-handed flagellar bundles are rotating counter-clockwise when seen from outside the cell and the bundles are closer to the agar surface than the top swarm/air surface. This chiral flow provides an avenue for long-range communication in the swarming colony. Furthermore, the observed chiral fluid flux of about $Q=40~(\mu$m$)^3/s$ suggest that bacterial flagella may be used to pump fluid near boundaries, as recently demonstrated experimentally \cite{gao2015using}.

%%%%%%%%
\section{Conclusion}

In conclusion, our proposed  model for the flows around bacterial swarms is  able to capture the observed chiral flow qualitatively and quantitatively and is  consistent with the rotation of the  flagella as the origin of the flow.  Flagella are sticking almost radially out of the swarm, suggesting that the flagella rotation (rotlet) is more important in creating the chiral flow along the swarm edge  than that induced by the  net forces (stokeslet) along the flagellar axes. This in turn suggests that the flagellar forces are   significantly contributing to the radial pumping of the fluid and the swarm expansion.  For the observed tangential flows in the clockwise direction one needs to have the counter-clockwise rotating left-handed flagella which are on average closer to the agar than the air/fluid interface, and a fit to our model  confirms this hypothesis, with an estimate of the height of the flagellar bundles  above the agar surface of $\hat{h}=h/H \approx 0.3-0.45$.  Moreover, our model estimates that the average distance between protruding flagella is $L\approx 5-8~\mu$m, just above one  average cell length.  This work provides a fundamental understanding  of the flows driven by flagella near   boundaries, and can be exploited in future studies to address force and torque generation in confined geometries. As an example, one could use our framework to quantify the flows and the mixing  induced by bacterial carpets in the limit of strong confinement  \cite{darnton2004moving,kim2007use}.

\section*{Acknowledgements}
We would like to thank Howard Berg for useful discussions. This work was funded in part by the EPSRC (JD) and the European Union through a Marie Curie CIG Grant (EL). 

\appendix
\section{Two parallel walls with one no-slip and one no-shear boundary conditions}
\label{A}
\begin{figure}[t]
\centering
\includegraphics[width=0.49\textwidth]{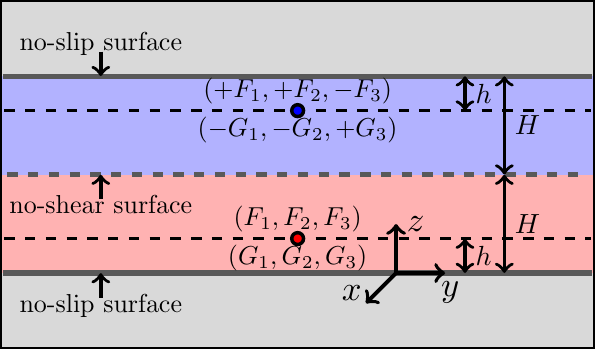}
\caption{Solution due to a point force/torque between two infinite parallel surfaces, a no-slip wall  at $z=0$ and a no-shear surface at  $z=H$. The solution is a superposition of two solutions for a point force/torque between parallel no-slip surfaces at  $z=0$ and  $z=2H$.}
\label{fig:14}
\end{figure}

Consider  two parallel walls, a no-slip wall at $z=0$ and a no-shear wall at $z=H$. Assume that there is a stokeslet $(F_1,F_2,F_3)$ and a rotlet $(G_1,G_2,G_3)$ located at $(x_0, y_0, h)$. 
The far-field solution will be mathematically equivalent to the case when we have two no-slip walls placed at $z=0$ and at $z=2H$, and two singularities: the first one is the original singularity placed at $z=h$ with $(F_1,F_2,F_3)$ and $(G_1, G_2, G_3)$, the second singularity is such that it gives no-shear
at $z=H$, i.e.~placed at $z=2H-h$ with $(+F_1,+F_2,-F_3)$ and $(-G_1,-G_2,+G_3)$, see Fig.~\ref{fig:14}.

Write $\hat{x}=kx$, $\hat{y}=ky$, $\hat{x}_0=kx_0$, $\hat{y}_0=ky_0$ and $z'=z/(2H)$, $h'=h/(2H)$ then the flow due to these two singularities is
\begin{align}
 u_i &= \frac{D_j k^2}{2\pi} \left(-\frac{\delta_{ij}}{\hat{r}^2}+\frac{2 \hat{r_i} \hat{r_j}}{\hat{r}^4}\right), \\
 u_3 &= 0,\\
 D_j &= D_j^{F}+D_j^{G},
\end{align}
where the source dipole strength due to two stokeslets is
\begin{equation}
\begin{split}
 D_j^{F} &= \frac{3}{\mu}\left[F_j (2H) h'\left(1-h'\right)z' \left(1-z'\right)+ \right. \\
 &\quad +\left. F_j (2H)\left(1-h'\right)h' z' \left(1-z'\right)\right]
 \end{split}
 \end{equation}
 and the source dipole strength due to two rotlets is 
 \begin{equation}
 \begin{split}
 D_j^{G} &= \frac{3}{\mu}\left \{ \epsilon_{3jm}G_m \left(\frac{1}{2}-h'\right)z' \left(1-z'\right) \right.\\
 &\quad +\left. \epsilon_{3jm} (-G_m) \left[\frac{1}{2}-(1-h')\right]  z' \left(1-z'\right) \right \}.
 \end{split}
  \end{equation}
  Simplify these expressions to get that
  \begin{align}
 D_j^{F}&= \frac{3}{\mu}\left[2F_j (2H) h'\left(1-h'\right)\right]  z' \left(1-z'\right),\\
 D_j^{G} &= \frac{3}{\mu}\left [ \epsilon_{3jm} 2G_m \left(\frac{1}{2}-h'\right)  \right]  z' \left(1-z'\right),
 \end{align}
and $i, j, m=1, 2$. 
Since we want to scale lengths with $H$ instead of $2H$, we write $\hat{h}=h/H$ and $\hat{z}=z/H$ then
  \begin{align}
 D_j^{F}&= \frac{3}{4\mu}\left[F_j H \hat{h} \left(2-\hat{h}\right)\right]  \hat{z} \left(2-\hat{z}\right),\\
 D_j^{G} &= \frac{3}{4\mu}\left [ \epsilon_{3jm} G_m \left(1-\hat{h}\right)  \right]  \hat{z} \left(2-\hat{z}\right),
 \end{align}
This solution is valid only in the far field, so $r \gg H$.
Notice that flows are half-parabolic in the direction perpendicular to walls, i.e.~proportional to $\hat{z}(2-\hat{z})$. The flows are in two dimensions and they are decaying as $1/r^2$ at large distances. 
Here again, the two-dimensional source dipole due to a stokeslet is in the direction of the force applied to the fluid, whereas the source dipole due to a rotlet is in the perpendicular
direction with respect to the applied torque. This time the sign of $D_j^G$ only depends on $G_m$ and not on the height $\hat{h}$. This means that the counter-clockwise rotating flagellar bundle will induce the tangential flow in the clockwise direction around the swarm due to the torque applied on the fluid.

\section{Fundamental solutions}
\label{B}

\begin{figure}[t]
\centering
\includegraphics[width=0.45\textwidth]{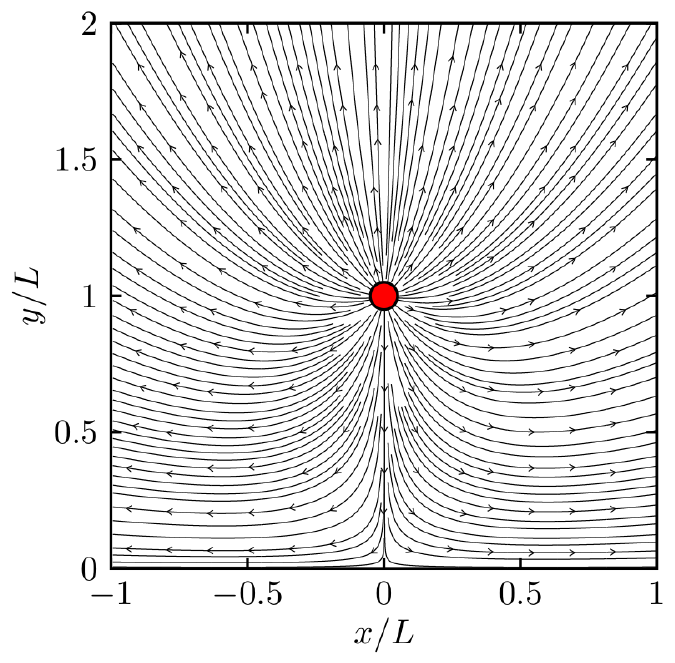}
\includegraphics[width=0.45\textwidth]{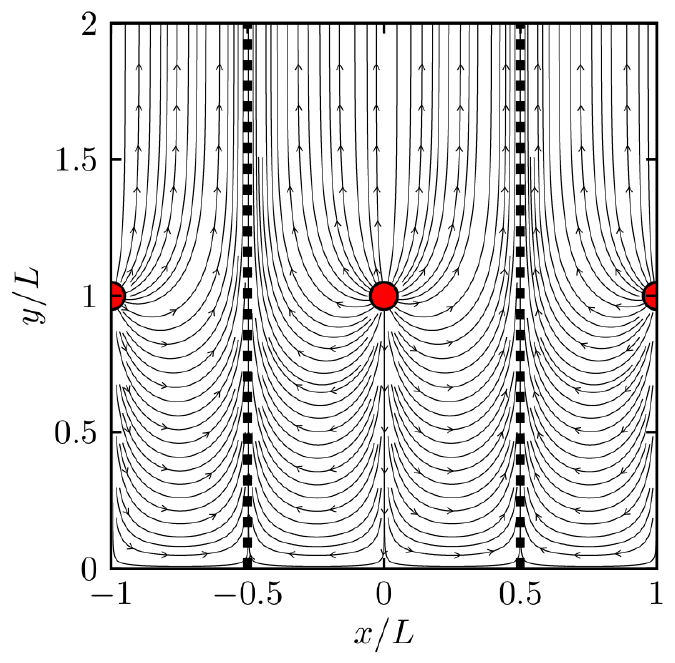}
\caption{Streamlines for a two-dimensional source above the no slip wall at $y=0$ (left) and for  a two-dimensional source above a no-slip wall with periodic boundary conditions (right).}
\label{fig:12}
\end{figure}

\subsection{A two-dimensional source  near a wall}
The flow due to a source in two dimensions of strength $M$ located at $(x_0,y_0)$ near a wall at $y=0$ is given by
\begin{eqnarray}
 u_i^S&=&\frac{M}{2\pi}\left[\frac{r_i}{r^2}+\frac{R_i}{R^2}-2\left(\frac{R_i}{R^2}-\frac{2 R_i R_2^2}{R^4}\right) \right. \\
 &&\left. +2h \left(\frac{\delta_{i2}}{R^2}-\frac{2 R_i R_2}{R^4}\right)\right], 
 \notag
\end{eqnarray}
where $\vec{r}=(x-x_0,y-y_0)$, $\vec{R}=(x-x_0,y+y_0)$ and $x_0,y_0$ are constants.
Writing its  components we have
\begin{align}
 u_1^S&=\frac{M}{2\pi}\left(\frac{r_1}{r^2}-\frac{R_1}{R^2}+\frac{4y R_1 R_2}{R^4}\right),\\
 u_2^S&=\frac{M}{2\pi}\left(\frac{r_2}{r^2}+\frac{R_2}{R^2}-\frac{2y}{R^2}+\frac{4y R_2^2}{R^4}\right),
\end{align}
with streamlines shown in Fig.~\ref{fig:12}.

\subsection{A source dipole parallel to a wall}
\begin{figure}[t]
\centering
\includegraphics[width=0.45\textwidth]{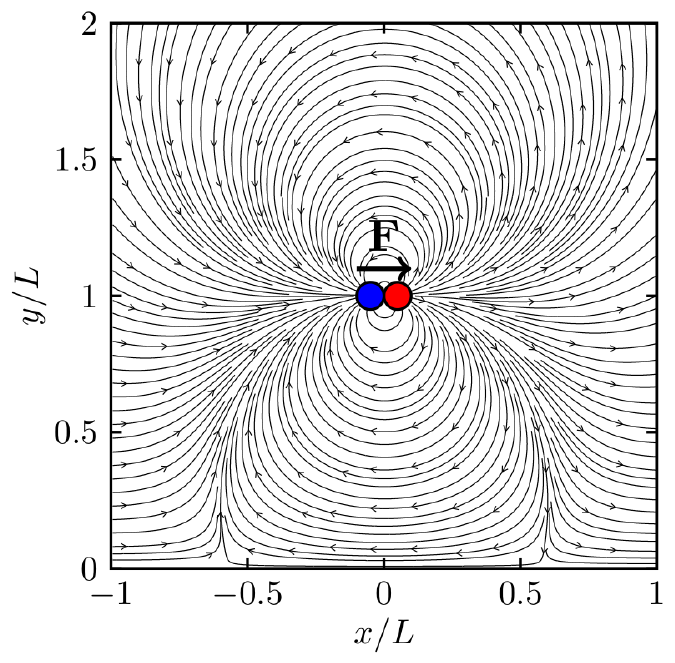}
\includegraphics[width=0.45\textwidth]{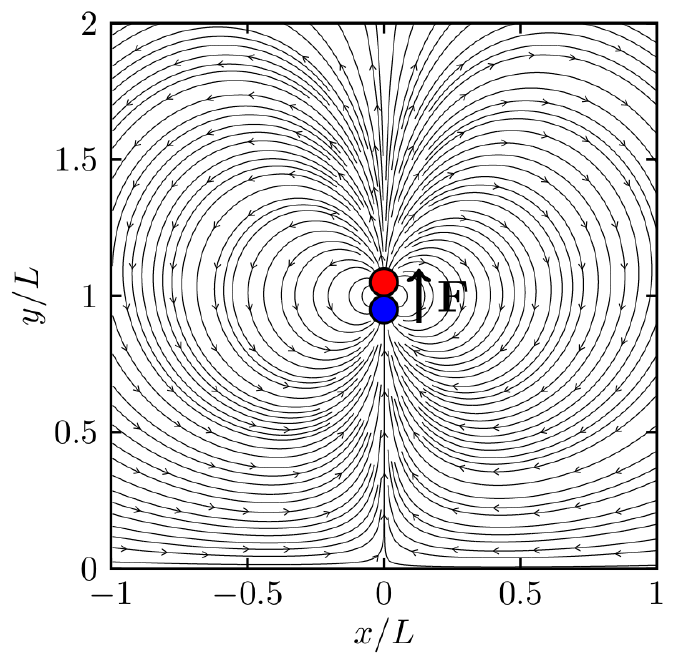}
\caption{Streamlines for two-dimensional source dipoles parallel (left) and perpendicular (right) to a  no-slip infinite wall at $y=0$. The source dipole is in the direction of the point force $F$.}
\label{fig:13}
\end{figure}

To find the solution for a source dipole parallel to a no-slip wall, we take the derivative of the source solution with respect to $x_0$
\begin{eqnarray}
 \vec{u}^{SD}(\vec{x};\vec{x_0};\vec{e_x})&=& l \frac{\partial}{\partial x_0} \vec{u}^S(\vec{x};\vec{x_0})\\
 &&=-l \frac{\partial}{\partial x} \vec{u}^S(\vec{x};\vec{x_0}),
\end{eqnarray}
where $l$ is the distance between the source and the sink, $\vec{x_0}=(x_0,y_0)$ is the location of singularity. Denote the strength of the source dipole $D=M l$, then the solution for the  dipole is  
 \begin{equation}
\begin{split}
 u_i^{SD}&(\vec{x};\vec{x_0};\vec{e_x})=-D\left[\left(\frac{\delta_{1i}}{r^2}-\frac{2r_1 r_i}{r^4}\right) \right.\\
 &-\left(\frac{\delta_{1i}}{R^2}-\frac{2R_1 R_i}{R^4}\right) \\
 &\left. +\frac{4\left(y R_2 \delta_{1i} - y_0 R_1 \delta_{2i} \right)}{R^4}-\frac{16 y R_1 R_2 R_i}{R^6}\right],
 \end{split}
\end{equation}
with streamlines  shown in Fig.~\ref{fig:13}

\subsection{A source dipole perpendicular to a wall}
Similarly we find a solution due to a source dipole perpendicular to a wall
\begin{equation}
 \vec{u}^{SD}(\vec{x};\vec{x_0};\vec{e_y})=l \frac{\partial}{\partial y_0} \vec{u}^S(\vec{x};\vec{x_0}),
\end{equation}
as
\begin{equation}
\begin{split}
 u_i^{SD}&(\vec{x};\vec{x_0};\vec{e_y})=-D\left[\left(\frac{\delta_{2i}}{r^2}-\frac{2r_2 r_i}{r^4}\right) \right.\\
 &-\left(\frac{\delta_{2i}}{R^2}-\frac{2R_2 R_i}{R^4}\right)\\
 &-\frac{4 R_2 (2y_0-R_2)\delta_{i2}+4 R_i (y_0-2 R_2)}{R^4}\\
 &\left.-\frac{16 y R_2^2 R_i}{R^6}\right],
 \end{split}
\end{equation}
with streamlines displayed in Fig.~\ref{fig:13}

\section{Periodic boundary conditions}
\label{C}
Using results  from Appendix \ref{B}, we  derive here the flow field due to a source dipole next to a no-slip wall with periodic boundary conditions.
\subsection{Source in the infinite fluid}
The flow solution for a source of strength $M$ in the infinite fluid is irrotational ,  $\vec{u}=\nabla \phi$, with potential given by
\begin{equation}
 \phi = \frac{M}{2 \pi} \ln{r},
\end{equation}
where $r=[(x-x_0)^2+(y-y_0)^2]^{1/2}$. 

Consider now  an array of sources located at $x_n=x_0+nL$, $y_n=y_0$ and $r_n=[(x-x_n)^2+(y-y_n)^2]^{1/2}$ for integer $n$. Then the solution due to 
array of sources is a linear superposition of the individual sources
\begin{equation}
 \phi = \sum_{n=-N}^{n=+N}\frac{M}{2 \pi} \ln{r_n}.
\end{equation}
Taking a limit as $N$ tends to infinity  one can formally sum up the potential and write \cite{lamb1932hydrodynamics, pozrikidis1992boundary}
\begin{equation}
 \phi = \frac{M}{2 \pi} A,
\end{equation}
where 
\begin{equation}
\begin{split}
 A&= \sum_{n=-\infty}^{+ \infty} \ln{r_n}\\
 &= \frac{1}{2} \ln{[\cosh{k(y-y_0)}-\cos{k(x-x_0)}]},
 \end{split}
\end{equation}
where $k={2\pi}/{L}$.
Therefore, the velocity field is given by
\begin{align}
\begin{split}
 u_1 &= \frac{M}{2 \pi} \frac{\partial A}{\partial x} \\
 &=\frac{Mk}{4 \pi} \frac{\sin{k(x-x_0)}}{\cosh{k(y-y_0)}-\cos{k(x-x_0)}},
 \end{split}
 \\
 \begin{split}
 u_2 &= \frac{M}{2 \pi} \frac{\partial A}{\partial y}\\
 &=\frac{Mk}{4 \pi} \frac{\sinh{k(y-y_0)}}{\cosh{k(y-y_0)}-\cos{k(x-x_0)}}\cdot
 \end{split}
\end{align}
\subsection{Source above the no-slip wall}
The solution for a single source above a no-slip wall in the components has components
\begin{align}
\begin{split}
 u_1 &=\frac{M}{2\pi}\left(\frac{r_1}{r^2}-\frac{R_1}{R^2}+\frac{4y R_1 R_2}{R^4}\right)\\
 &=\frac{M}{2\pi}\left(\frac{\partial}{\partial x}\ln{r}-\frac{\partial}{\partial x}\ln{R}-2y\frac{\partial^2}{\partial x \partial y}\ln{R}\right),
 \end{split}\\
 \begin{split}
 u_2 &=\frac{M}{2\pi}\left(\frac{r_2}{r^2}+\frac{R_2}{R^2}-\frac{2y}{R^2}+\frac{4y R_2^2}{R^4}\right)\\
 &=\frac{M}{2\pi}\left(\frac{\partial}{\partial y}\ln{r}+\frac{\partial}{\partial y}\ln{R}-2y\frac{\partial^2}{\partial y^2}\ln{R}\right),
 \end{split}
\end{align}
where 
$\vec{r}=(x-x_0,y-y_0)$, $\vec{R}=(x-x_0,y+y_0)$.
Writing
\begin{equation}
\begin{split}
 B&= \sum_{n=-\infty}^{+ \infty} \ln{R_n}\\
 &= \frac{1}{2} \ln{[\cosh{k(y+y_0)}-\cos{k(x-x_0)}]},
 \end{split}
\end{equation}
then expressing the single source solution in derivatives of $\ln{r}$ and $\ln{R}$ we can simply obtain the periodic solution, namely
\begin{align}
 u_1^{S}&=\frac{Mk}{2\pi}\left(\frac{\partial A}{\partial \hat{x}} - \frac{\partial B}{\partial \hat{x}} - 2 \hat{y} \frac{\partial^2 B}{\partial \hat{x} \partial \hat{y}}\right),\\
 u_2^{S}&=\frac{Mk}{2\pi}\left(\frac{\partial A}{\partial \hat{y}} + \frac{\partial B}{\partial \hat{y}} - 2 \hat{y} \frac{\partial^2 B}{\partial \hat{y}^2}\right),
\end{align}
with streamlines that  are shown in Fig.~\ref{fig:12}.

\subsection{Source dipole above the no-slip wall}
The solution for the source dipole $\vec{u}^{SD}$ above the no-slip wall can be easily obtained by taking derivatives of the solution for the source $\vec{u}^{S}$
\begin{equation}
 \vec{u}^{SD}= l_1\frac{\partial \vec{u}^{S}}{\partial x_0}+l_2\frac{\partial \vec{u}^{S}}{\partial y_0},
\end{equation}
where $D_1=M l_1$ and $D_2=M l_2$ are the source dipole strengths in the parallel and the perpendicular to the wall directions respectively. The flow  components are
\begin{align}
\begin{split}
 u_1^{SD}&= \frac{D_1 k^2}{2 \pi} \left(\frac{\partial^2 A}{\partial \hat{y}^2} -\frac{\partial^2 B}{\partial \hat{y}^2} - 2\hat{y} \frac{\partial^3 B}{\partial \hat{y}^3} \right)\\
 &+\frac{D_2 k^2}{2 \pi} \left(-\frac{\partial^2 A}{\partial \hat{x}\partial \hat{y}} - \frac{\partial^2 B}{\partial \hat{x}\partial \hat{y}} - 2\hat{y} \frac{\partial^3 B}{\partial \hat{x}\partial \hat{y}^2}\right),
 \end{split}
 \\
 \begin{split}
 u_2^{SD}&= \frac{D_1 k^2}{2 \pi} \left(-\frac{\partial^2 A}{\partial \hat{x}\partial \hat{y}} - \frac{\partial^2 B}{\partial \hat{x}\partial \hat{y}} + 2\hat{y} \frac{\partial^3 B}{\partial \hat{x}\partial \hat{y}^2}\right)\\
 &+\frac{D_2 k^2}{2 \pi} \left(-\frac{\partial^2 A}{\partial \hat{y}^2} +\frac{\partial^2 B}{\partial \hat{y}^2} - 2\hat{y} \frac{\partial^3 B}{\partial \hat{y}^3} \right).
 \end{split}
\end{align}
Note that the functions $A$ and $B$ are harmonic, a fact we used in $u_1^{SD}$ to write  $x$ derivatives in terms of $y$ derivatives.
One needs  derivatives of $A$ and  $B$ in order to compute the flow, and they are
\begin{align}
 \frac{\partial A}{\partial \hat{x}}&= \frac{\sin{(\hat{x}-\hat{x}_0)}}{2[\cosh{(\hat{y}-\hat{y}_0)}-\cos{(\hat{x}-\hat{x}_0)}]},\\
 \frac{\partial A}{\partial \hat{y}}&=\frac{\sinh{(\hat{y}-\hat{y}_0)}}{2[\cosh{(\hat{y}-\hat{y}_0)}-\cos{(\hat{x}-\hat{x}_0)}]},\\
  \frac{\partial^2 A}{\partial \hat{x}\partial \hat{y}}&=-\frac{\sin{(\hat{x}-\hat{x}_0)}\sinh{(\hat{y}-\hat{y}_0)}}{2[\cosh{(\hat{y}-\hat{y}_0)}-\cos{(\hat{x}-\hat{x}_0)}]^2},\\
 \frac{\partial^2 A}{\partial \hat{y}^2} &=\frac{1-\cos{(\hat{x}-\hat{x}_0)}\cosh{(\hat{y}-\hat{y}_0)}}{2[\cosh{(\hat{y}-\hat{y}_0)}-\cos{(\hat{x}-\hat{x}_0)}]^2},\\
\begin{split}
 \frac{\partial^3 A}{\partial \hat{y}^3} &=\frac{\cos{(\hat{x}-\hat{x}_0)}\cosh{(\hat{y}-\hat{y}_0)}\sinh{(\hat{y}-\hat{y}_0)}}{2[\cosh{(\hat{y}-\hat{y}_0)}-\cos{(\hat{x}-\hat{x}_0)}]^3}\\
 &-\frac{[1+\sin^2{(\hat{x}-\hat{x}_0)}]\sinh{(\hat{y}-\hat{y}_0)}}{2[\cosh{(\hat{y}-\hat{y}_0)}-\cos{(\hat{x}-\hat{x}_0)}]^3},
 \end{split}
 \\
 \begin{split}
  \frac{\partial^3 A}{\partial \hat{x}\partial \hat{y}^2} &=\frac{\cos{(\hat{x}-\hat{x}_0)}\cosh{(\hat{y}-\hat{y}_0)}\sin{(\hat{x}-\hat{x}_0)}}{2[\cosh{(\hat{y}-\hat{y}_0)}-\cos{(\hat{x}-\hat{x}_0)}]^3}\\
  &-\frac{[1-\sinh^2{(\hat{y}-\hat{y}_0)}]\sin{(\hat{x}-\hat{x}_0)}}{2[\cosh{(\hat{y}-\hat{y}_0)}-\cos{(\hat{x}-\hat{x}_0)}]^3}
  \end{split}
\end{align}
with similar  expressions for $B$  obtained by changing $y_0$ to $-y_0$.
For the simple case, take $\hat{x}-\hat{x}_0=\pi$ then $\cos(\pi)=-1$ and $\sin(\pi)=0$. This gives
\begin{align}
 \frac{\partial^2 A}{\partial \hat{y}^2} &=\frac{1}{2[1+\cosh{(\hat{y}-\hat{y}_0)]}},\\
 \frac{\partial^2 B}{\partial \hat{y}^2} &=\frac{1}{2[1+\cosh{(\hat{y}+\hat{y}_0)]}},\\
 \frac{\partial^3 B}{\partial \hat{y}^3} &=-\frac{\sinh{(\hat{y}+\hat{y}_0)}}{2[1+\cosh{(\hat{y}+\hat{y}_0)}]^2}.
\end{align}
The tangential velocity profile is given by
\begin{equation}\label{tangential}
\begin{split}
 u_1&=\frac{D_1 k^2}{4 \pi} \left(\frac{1}{1+\cosh{(\hat{y}-\hat{y}_0)}} \right. \\
 &\left. -\frac{1}{1+\cosh{(\hat{y}+\hat{y}_0)}} + \frac{2\hat{y} \sinh{(\hat{y}+\hat{y}_0)}} {[1+\cosh{(\hat{y}+\hat{y}_0)}]^2} \right),
 \end{split}
\end{equation}
and the radial velocity profile 
\begin{equation}
\begin{split}
 u_2&=\frac{D_2 k^2}{4 \pi} \left(-\frac{1}{1+\cosh{(\hat{y}-\hat{y}_0)}} \right. \\
 &\left. +\frac{1}{1+\cosh{(\hat{y}+\hat{y}_0)}} + \frac{2\hat{y} \sinh{(\hat{y}+\hat{y}_0)}} {[1+\cosh{(\hat{y}+\hat{y}_0)}]^2} \right).
 \end{split}
\end{equation}
The streamlines for the periodic dipole near a wall are  shown in   Fig.~\ref{fig:4}. The  dipole perpendicular to the no-slip boundary generates no net flow along the boundary by symmetry, but the  dipole in the direction along the wall generates a net flow rate, which can be calculated analytically.

\subsection{Flow rates}
The flow rates due to a source dipole near a wall with periodic boundary conditions  along the swarm edge, $q_1$, and perpendicular to it,  $q_2$, are defined as
\begin{align}
q_1&=\int_0^{+\infty}u_1 dy,\\
q_2&=\int_{-\infty}^{+\infty}u_2 dx.
\end{align}
Since there is a wall at $y=0$ we have $q_2=0$. For
 the  flow rate along the swarm edge,  only the source dipole parallel to the wall contributes to the net flux since
\begin{equation}
\begin{split}
q_1&=\int_0^{+\infty} u_1^{SD}dy\\
&=\int_0^{+\infty} \frac{D_1 k^2}{2 \pi} \left(\frac{\partial^2 A}{\partial \hat{y}^2} -\frac{\partial^2 B}{\partial \hat{y}^2} - 2\hat{y} \frac{\partial^3 B}{\partial \hat{y}^3} \right) dy\\
&=\frac{D_1 k}{2 \pi}=\frac{D_1}{L}.
\end{split}
\end{equation}
Note that this is an exact result. 

\subsection{Asymptotic expansion}
Suppose for simplicity that $\hat{x}-\hat{x}_0=\pi$. We want to investigate the flow far away from the singularity $y_0$, i.e.~in the limit $\hat{y}-\hat{y}_0 \gg 1$. This means that $\hat{y}+\hat{y}_0 \gg 1$ since $\hat{y}_0 > 0$. The tangential flow profile in this limit using Eq.~\eqref{tangential} is
\begin{equation}
\begin{split}
 u_1&=\frac{D_1 k^2}{2 \pi} \left(e^{-(\hat{y}-\hat{y}_0)}- e^{-(\hat{y}+\hat{y}_0)}+2\hat{y}e^{-(\hat{y}+\hat{y}_0)} \right)\\
 &+\mathcal{O}\left(e^{-2(\hat{y}-\hat{y}_0)},~ e^{-2(\hat{y}+\hat{y}_0)},~\hat{y}e^{-2(\hat{y}+\hat{y}_0)}\right),
 \end{split}
\end{equation}
which can be rewritten to a leading order as 
\begin{equation}
\begin{split}
 u_1&=\frac{D_1 k^2 e^{-\hat{y}}}{\pi} \left(\sinh{\hat{y}_0}+\hat{y}e^{-\hat{y}_0}\right)
 \end{split}
\end{equation}
Now there are two limits: if $\hat{y}_0 \ll 1$ or equivalently if $y_0 \ll L$ then we have
\begin{equation}
\begin{split}
 u_1&=\frac{D_1 k^2 e^{-\hat{y}}}{\pi}\hat{y}
 \end{split}
\end{equation}
in contrast if $\hat{y}_0 \gg 1$ or equivalently if $y_0 \gg L$ then we obtain 
\begin{equation}
\begin{split}
 u_1&=\frac{D_1 k^2 e^{-(\hat{y}-\hat{y}_0)}}{2\pi}
 \end{split}
\end{equation}
\bibliographystyle{unsrt}
\bibliography{justas}

\end{document}